\magnification=\magstep1
\tolerance=500
\magnification=\magstep1
\tolerance=500
\rightline{TAUP 2775-04}
\rightline{20 July, 2004}
\bigskip
\centerline{\bf Classical Gravity as an Eikonal Approximation}
\centerline{\bf to a }
\centerline{\bf Manifestily Lorentz Covariant Quantum Theory}
\centerline {\bf with}
\centerline{\bf Brownian Interpretation}
\bigskip
\centerline{Lawrence P. Horwitz and Ori Oron}
\centerline{School of Physics}
\centerline{Tel Aviv University}
\centerline{Ramat Aviv 69978, Israel}
\bigskip
\noindent
{\it Abstract:\/}
\par We discuss in this Chapter a series of theoretical developments
which motivate the introduction of a quantum evolution equation for
which the eikonal approximation results in the geodesics of a four
dimensional manifold.  This geodesic motion can be put into
correpondence with general relativity.  The well-known problem of the
self-interaction of a relativistic charged particle is studied from
the point of view of a manifestly Lorentz covariant classical theory
admitting the particle mass as a dynamical variable, i.e., the theory
is intrinsically off-shell.  The evolution parameter, $\tau$, is an
invariant parameter that can be identified with Newton's time
(sometimes called, as in Schwinger's  formulation, ``proper
time''). Gauge invariance requires the definition of five gauge
fields; the fifth field has for its source the matter
density. Together with the results of Gupta and Padmanabhan, showing a
connection between the radiation reaction force and geometry, this
structure motivates an investigation into the connection between these
dynamical equations and gravity. We show that the fifth gauge field
can indeed be absorbed into a conformal metric in the kinematic terms, which
then results in a geodesic equation generated by the conformal metric
and the standard Lorentz force.  We then go on to show that the
generalized radiation field passing through an optical medium with
non-trivial dielectric tensor results in an analog gravity for the
eikonal approximation for an arbitrary metric.  A mathematically
simpler system for
which the eikonal approximation provides the geodesic motion on a four
dimensional pseudo-Riemannian manifold is that of the
Stueckelberg-Schr\"odinger equation with a spacetime dependent tensor
 $g_{\mu\nu}$
(of the form of the Einstein metric tensor), somewhat analogously to a
gauge field, coupling to the kinetic
terms.  This theory can be realized as a quantum theory in a flat
spacetime, obeying the rules of the standard quantum theory in Lorentz
covariant form.  Since the geodesics predicted by the eikonal
approximation, with appropriate choice of $g_{\mu\nu}$, can
be those of general relativity, this theory provides a quantum theory
which underlies classical gravitation, and coincides with it in this
classical ray approximation. This result is the principal content of
this work.
\par In order to understand the possible origin of the structure of
this form of the Stueckelberg-Schr\"odinger equation, we appeal
 to the approach of Nelson in constructing a Schr\"odinger equation
 from the properties of Brownian motion.  Extending the notion of
 Brownian motion to spacetime in a covariant way, we show that such an
 equation follows from correlations between
 spacetime dimensions in the stochastic process.

\vfill
\break
\bigskip
\noindent{\bf I. Introduction}
\par We shall deal with the problem of constructing a quantum
evolution equation which describes, in the eikonal approximation, 
the geodesics of general relativity, on essentially three levels
which include the basic motivations arising from the self-interaction
problem of a relativistic charged particle in the framework of a
generalized electromagetic theory, an analysis of the
conformal structure that is associated with this theory, and the
discussion of the eikonal approximation of wave equations that lead to
our basic result.  To understand the structure of the realtivistically
covariant quantum equation, we study a relativistic generalization of 
Brownian motion, and
follow Nelson's construction of the Schr\"odinger equation [1] for which
local spacetime correlations result in a modification of the
kinematic terms by introducing coupling to a Lorentz tensor
$g_{\mu\nu}$ somewhat analogous to a gauge field.  
 \par  It is quite remarkable that
Gupta and Padmanabhan [2], using essentially
geometrical arguments (solving the static problem in the frame of the
 accelerating particle with
  a curved background metric), have shown that the description of
  the motion of an accelerating charged particle {\it must} include the
 radiation terms of the Abraham-Lorentz-Dirac equation [3], 
$$ m {d^2 x^\mu \over ds^2} = F^\mu_\nu {dx^\mu \over ds} +
\Gamma^\mu, \eqno(1.1)$$
where $m$ is the electron mass, including electromagnetic correction,
 $s$ is the proper time
along the trajectory $x^\mu(s)$ in spacetime, $F^\mu_\nu$ is the
covariant form of the electromagnetic force tensor, $e$ is the
electron charge, and 
$$ \Gamma^\mu = {2 \over 3}{e^2 \over c^3} \bigl( {d^3 x^\mu \over
ds^3} - {d^2x^\nu \over ds^2}{d^2 x_\nu \over ds^2} {dx^\mu \over ds}
\bigl)
\eqno(1.2)$$
\par Here, the indices $\mu, \nu$, running over $0,1,2,3$, label the
spacetime variables that represent the action of the Lorentz group;
 the index raising and lowering Lorentz invariant tensor $\eta_\mu\nu$ is of
 the form $diag(-1,+1,+1,+1)$.  The expression for $\Gamma^\mu$ was
 originally found by Abraham in 1905 [3], shortly after the discovery of
 special relativity, and is known as the Abraham four-vector of
 radiation reaction. Dirac's derivation [3] was based on a direct
 application of the Green's functions for the Maxwell fields,
 obtaining the form $(1.1)$.
\par  Recognizing
that the electron's acceleration precludes the use of a sequence of
``instantaneous'' inertial frames to describe the action of the
forces on the electron [4], they carry out a Fermi-Walker
transformation [5], going
to an accelerating frame (assuming constant acceleration) in which the
electron is actually inertial, and there solve the Coulomb problem in
the curved coordinates provided by the Fermi-Walker transformation.
Transforming back to laboratory coordinates, they find the
Abraham-Lorentz-Dirac equation without the direct use of the Maxwell
Green's functions for the radiation field. This result, suggesting the
relevance of curvature in the spacetime manifold, such as that
generated by sources in general relativity, along with other,
more elementary manifestations of mass renormalization (such as the
contribution to the mass due to electromagnetic interactions and the
identification of the Green's function singularity contribution with
part of the electron mass), carries an implication that the electron mass        may
play an important {\it dynamical} role.
\par Stueckelberg, in 1941 [6], proposed a manifestly covariant form of
classical and quantum mechanics in which space and time become
dynamical observables.  They are therefore represented in quantum
theory by operators on a Hilbert space on square integrable functions
in space and time.  The dynamical development of the state is
controlled by an invariant parameter $\tau$, which one might call the
world time, coinciding with the time on the (on mass shell)freely
falling clocks of general relativity.  Stueckelberg [6] started his
analysis by considering a classical world-line, and argued that under
the action of forces, the world line would not be straight, and in
fact could be curved back in time.  He identified the branch of the
curve running backward in time with the antiparticle, a view taken
also by Feynman in his perturbative formulation of quantum
electrodynamics in 1948 [6].  Realizing that such a curve could not be
parametrized by $t$ (for some values of $t$ there are two values of
the space variables), Stueckelberg introduced the parameter $\tau$ along the
trajectory. 
\par This parameter is not necessarily identical to proper
time, even for inertial motion for which proper time is a meaningful
concept. Stueckelberg postulated the existence of an invariant
``Hamiltonian'' $K$, which would generate Hamilton equations for the
canonical variables $x^\mu$ and $p^\mu$ of the form
$$ {\dot x}^\mu = {\partial K \over \partial p_\mu} \eqno(1.3)$$
and 
$$ {\dot p}^\mu =  -{\partial K \over \partial x_\mu}, \eqno(1.4)$$
where the dot indicates differentiation with respect to $\tau$.
Taking, for example, the model
$$ K_0 = {p^\mu p_\mu \over 2M}, \eqno(1.5)$$
we see that the Hamilton equations imply that
$$ {\dot x}^\mu = {p^\mu \over M} \eqno(1.6)$$
It then follows that
$$ {d{\bf x} \over dt} = {{\bf p} \over E}, \eqno(1.7)$$
where $p^0 \equiv E$, where we set the velocity of light $c=1$; this is the 
correct definition for the velocity of a free relativistic particle.
It follows, moreover, that
$$ {\dot x}^\mu  {\dot x}_\mu = {dx^\mu dx_\mu \over d\tau^2} = {p^\mu p_\mu 
\over M^2} \eqno(1.8)$$
With our choice of metric, $dx^\mu dx_\mu = -ds^2$, and $p^\mu p_\mu = -m^2$,
where $m$ is the classical experimentally measured mass of the
particle (at a given instant of $\tau$).  We 
see from this that 
$$ {ds^2 \over d\tau^2} = {m^2 \over M^2}, \eqno(1.9)$$
and hence the proper time is not identical to the evolution parameter
$\tau$.
 In 
the case that $m^2 = M^2$, it follows that  $ds = d\tau$, and we say that the 
particle is ``on shell''.
\par For example, in the case of an external potential $V(x)$, where
we write
 $x \equiv x^\mu$, the Hamiltonian becomes
$$ K = {p^\mu p_\mu \over 2M} + V(x) \eqno(1.10)$$
 so that, since $K$ is a constant of the motion, $m^2$ varies from point to point 
with the variations of $V(x)$. It is important to recognize from this discussion 
that the observable particle mass depends on the {\it state} of the system (in the 
quantum theory, the expectation value of the operator $p^\mu p_\mu$
provides
 the 
expected value of the mass squared).
 \par One may see, alternatively, that phenomenologically the mass of
a nucleon, 
such as the neutron, clearly depends on the state of the system.  The free 
neutron is not stable, but decays spontaneously into a proton, electron and 
antineutrino, since it is heavier than the proton.  However, bound in
a nucleus, 
it may be stable (in the nucleus, the proton may decay into neutron,
 positron and 
neutrino, since the proton may be sufficiently heavier than the neutron).  The 
mass of the bound electron (in interaction with the electromagnetic field), as 
computed in quantum electrodynamics, is different from that of the
 free electron, 
and the difference contributes to the Lamb shift.  This implies that, if one 
wishes to construct a covariant quantum theory, the variables $E$ (energy) and 
$\bf{ p}$ should be independent, and not constrained by the relation
 $E^2 = {\bf 
p}^2 + m^2,$ where $m$ is a fixed constant.  This relation
implies, moreover, that $m^2$ is a dynamical variable.  It then 
follows, quantum mechanically,  through the Fourier relation between the 
energy-momentum representation of a wave function and the spacetime 
representation, that the variable $t$, along with the variable ${\bf x}$ is a 
dynamical variable. Classically, $t$ and ${\bf x}$ are recognized as
 variables of 
the phase space through the Hamilton equations.
\par Since, in nature, particles appear
with fairly sharp mass values (not necessarily with zero spread), we may
assume the existence of some mechanism which will drive  the
particle's mass back to its original mass-shell value (after the
source responsible for the mass change ceases to act) so that the
particle's mass shell is defined.   We shall not take such a
mechanism into account explicitly here in developing the dynamical
equations. We shall assume that this mechanism is working, and it is a
relatively smooth function (for example, a minimum in free energy
which is broad enough for our off-shell driving force to work
 fairly freely)\footnote{${}^1$}{
  A relativistic
Lee model has been worked out which describes a physical mass shell
as a resonance, and therefore a stability point on the spectrum [7],
 but at this
point it is not clear to us how this mechanism works in general.  It
has been suggested by T. Jordan [personal communication] that the
definition of the physical mass shell could follow from the
interaction of the particle with fields (a type of
``self-interaction''); this mechanism could provide for perhaps more
than one mass state for a particle, such as the electron and muon and
the various types of neutrinos, but no detailed model has been so far
studied.}.
\par In an application of statistical mechanics to this theory [8], it has
been found that a high temperature phase transition can be responsible
for the restriction of the particle's mass (on the average, in
equilibrium). In the
classical theory, the non-linear equations induced by radiation
reaction may have a similar effect [9].
\par  The 
Stueckelberg formulation implies the existence of a fifth ``electromagnetic'' 
potential, through the requirement of gauge invariance, and there is a 
generalized Lorentz force which contains a term that drives the particle 
off-shell, whereas the terms corresponding to the electric and
magnetic parts
 of the usual Maxwell fields do not (for the nonrelativistic case, the
electric
 field 
may change the energy of a charged particle, but not the magnetic field; the 
electromagnetic field tensor in our case is analogous to the usual magnetic
field, and the new field strengths, derived from the $\tau$ dependence
of the fields and the additional gauge field, are analogous to the usual
electric field, as we 
shall see).  The second quantization of this gauge theory has been
carried put as well [10].
\par In the following, we give the structure of the field equations,
and show that 
the standard Maxwell theory is properly contained in this more general
 framework.  
Applying the Green's functions to the current source provided by the
 relativistic 
particle, and the generalized Lorentz force, we obtain equations of
motion
 for the 
relativistic particle which is, in general off-shell.  As in Dirac's result, 
these equations are of third order in the evolution parameter, and
therefore are highly unstable.  However, the equations are very
nonlinear, and give rise to chaotic behavior [9].
\par Our results exhibit what appears to be a strange attractor in the
phase space of the 
autonomous equation for the off-mass shell deviation. This attractor may 
stabilize the electron's mass in some neighborhood. We conjecture that
it stabilizes the orbits macroscopically as well, but a detailed
analysis awaits the application of more powerful computing facilities
 and procedures.
\par  We then show that the
fifth (scalar) field can be eliminated through the introduction of a
conformal metric on the spacetime manifold [11].  The geodesic equation
associated with this metric coincides with the Lorentz force, and is
therefore dynamically equivalent.  Since the generalized Maxwell
equations for the five dimensional fields provide an equation relating
the  fifth field with the spacetime density of events, one can derive
 the spacetime event density associated with the Friedmann-Robertson-Walker
solution of the Einstein equations.  The resulting density, in the
conformal coordinate space, is isotropic and homogeneous,
decreasing as the square of the Robertson-Walker scale factor.  Using
the Einstein equations, one see that both for the static and matter
dominated models, the conformal time slice in which the events which
generate the world lines are contained becomes progressively thinner as
the inverse square of the scale factor, establishing a simple
correspondence between the configurations predicted by the underlying
Friedmann-Robertson-Walker dynamical model and the 
configurations in the conformal coordinates.
\par The conformal metric is not, however, even locally equivalent to
a Schwarzschild metric.  To achieve a more general framework for
achieving an underlying model for gravity, we study the eikonal
approximation of the (generalized) electromagnetic equations in a
medium with non-trivial dielectric tensor.
\par It has been known for many years that the Hamilton-Jacobi
equation of classical mechanics
defines a function which appears to be the eikonal of a wave equation,
and therefore that classical mechanics appears to be a ray
approximation to some wave theory [12]. The propagation of rays of
waves in inhomogeneous media appears, from this point of view (as a
result of the application of Fermat's principle), to correspond to
geodesic motion in a metric derived from the properties of the
medium [13].   This geometrical
interpretation has been exploited recently by several authors to
construct models which exhibit three dimensional analogs of general
relativity by studying the wave equations of light in an inhomogeneous
medium [14], and, to achieve four dimensional analogs,
 sound waves and electromagnetic propagation in
 inhomogeneously moving materials [15].  Visser {\it et al\/} [16]
have pointed out that condensed matter systems such as acoustics in flowing
fluids, light in moving dielectrics, and quasiparticles in a moving
superfluid can be used to mimic kinematical aspects of general
relativity. Leonhardt and Piwnicki [17] and Lorenci and Klippert [18],
for example, have discussed the case of electromagnetic propagation in
moving media.   In order to  achieve four dimensional geodesic flows, it has
been necessary to introduce a motion of the medium. 
\footnote{${}^2$}{We remark that Obukhov and Hehl [19] have shown that a conformal class
of metrics for
spacetime can be derived by imposing constrained linear constitutive
relations between the electromagnetic fields $(E,B)$ and the
excitations $(D,H)$, using Urbantke's formulas [20], developed to
define locally integrable parallel transport orbits in Yang-Mills
theories (on tangent 2-plane elements on which the Yang-Mills
curvature vanishes).}
  There is considerable interest in extending these
analog models for the kinematical aspects of gravity to include
dynamical aspects, {\it i.e.} considering gravity as an emergent 
phenomenon [16].
\par The manifestly covariant classical and quantum mechanics
introduced by \hfil\break
Stueckelberg [6] in 1941 has the structure of the
Hamilton dynamics with the Euclidean three dimensional space replaced
by four-dimensional Minkowski space (since all four of the components
of energy-momentum are kinematically independent, the theory is
intrinsically ``off-shell'').  This theory leads to five dimensional
wave equations for the associated gauge fields (in fact, for wave
phenomena in general, such as acoustic or hydrodynamic waves), which
in the eikonal approximation in the presence of
an inhomogeneous medium, provides a basis
for geodesic motion in four dimensional spacetime.
\par This theory has been used 
to account for the known bound state spectra of the (spinless)
two-body problem in potential theories formulated in a manifestly
covariant way [21] (as well as to study the classical relativistic
Kepler orbits [6,22]).  As we have remarked above, in order for the
 Stueckelberg-Schr\"odinger
equation of the quantum form of this theory to be gauge invariant, it
is necessary to introduce a fifth gauge field, compensating for the
derivative of the invariant evolution parameter $\tau$ [10].
Generalized gauge invariant field strengths, $f_{q,p}$, with $q,p
= 0,1,2,3,5$ occurring in the Lagrangian to second order generate
second order field equations analogous to the usual Maxwell equations,
with source given by the four-vector mattter field current and an
additional Lorentz scalar density. 
 Taking the Fourier transform of
these equations over the invariant parameter, as we demonstrate below,
one sees that the zero
frequency component (zero mode) of the equations coincides with the
standard Maxwell theory. The field equations
go over in the same way; the equation for the fifth field decouples.
The Maxwell theory is therefore properly contained in the
five-dimensional generalization as the zero frequency component. In the quantum
case the four-currents are given by bilinears in the wave
function containing first derivatives and the fifth source is the
(Lorentz) scalar probability density. The symmetry of the
homogeneous equations, which can be $O(3,2)$ or $O(4,1)$, depending on
the sign chosen for raising and lowering the fifth index, is therefore not
realized in the inhomogeneous equations.  Hence, without augmenting
the symmetry of the {\it matter} fields beyond $O(3,1)$, the
fifth field, whose source in the Maxwell-like equations is the
probability density (or, classically,the matter density) in
spacetime, can play a special role. There appears to be no kinematic
basis for choosing one or the other of these signatures; atomic
radiative decay, for example, contains points in phase space (for
radiation of off-shell photons) for either type.  We note, however,
that the homogeneous equations corresponding to the $O(4,1)$ signature
appear, under Fourier transform of the $\tau$ variable, as 
Klein-Gordon type wave equations with positive mass-squared (physical
particles) , but for the $O(3,2)$ choice of signature, these equations
have the ``wrong'' sign (tachyonic) for interpreting the additive term
as a mass-squared.  We do not find an objective criterion for choosing
one or the other of these possibilities, and therefore leave open the
question of a definitive choice of the signature for the five
dimensional radiation field at this stage.  It is possible that both
may play a dynamical role.
\par The structure of the gauge theory obtained from the
non-relativistic Schr\"odinger equation is precisely analogous.  The
four dimensional gauge invariant field strengths obey second order
equations, for which the sources are the vector currents and the scalar
probability density (or, classically, the matter density in three
dimensional space). No linear transformation can connect the
Schr\"odinger probability density with the vector currents, so that
the scalar density and the fourth gauge field can play a special role.
\par  Since the eikonal approximation naturally lowers the dimension
of the differential equations describing the fields by one, the
eikonal approximation to the five-dimensional field equations results
in four dimensional differential equations.
 \par  In the presence of a
non-trivial dielectric structure of the medium, the four dimensional
field equations resulting from  the eikonal approximation can describe
 geodesic motion in four dimensional
spacetime without the necessity of adding motion to the medium. We
emphasize that the underlying manifold, on which the fields are
defined, is a flat Cartesian space, but that the
dynamically induced trajectories move on
the geodesics of a pseudo-Riemannian manifold. This
result forms our basic motivation for studying a generalized
dynamics of Stueckelberg [6] form in this context. 
\par In Section 5, we study  
the eikonal structure of waves in a 5D inhomogeneous medium, in which
 Minkowski spacetime is embedded, both for second order wave equations
 and a wave equation of Schr\"odinger type, and show that the resulting rays 
have a precise analog with the results of Kline and Kay [13], but in a
 4D spacetime manifold with a pseudo-Riemannian structure. Kline and
Kay [13] show that the rays are geodesic in the metric associated with
the anisotropic inhomogeneous medium.  As for the case of Kline and
Kay [13], it follows from the existence of a Hamiltonian that the
corresponding Lagrangian obeys an extremum condition which describes
the rays as geodesics.  We show
that there is mass flow along these rays, and that
the flow is controlled by generating functions of Hamiltonian type,
establishing a relation between geodesic flow and a particle mechanics
 of symplectic form. 
 We stress that this
construction does not necessary imply that the equation is written in a
curved spacetime, as for the wave equations of Birrell and Davies [23],
for example.  The quantum theory associated with this equation is
defined on a flat Minkowski spacetime; the wave functions satisfy
Euclidean normalization conditons, and four-momentum remains the
generator of spacetime translations.   It is only in the eikonal
approximation that the rays emerge as geodesics of a curved space.
\par  This result is the principal content of this
work.
\par We close our study here by considering a basis for the structure
of the Steuckelberg-Schr\"odinger equation with a second rank Lorentz
tensor applied to the kinematic terms.
\par The ideas of stochastic processes originated in the second half of the
 19th century in thermodynamics, through the manifestation of the
 kinetic theory of gases. In 1905 A. Einstein [24] in his
 paper on
 Brownian
 motion provided a decisive breakthrough in the understanding of the
 phenomena. Moreover, it was a proof convincing physicists of the
 reality of atoms and molecules,  the motivation for Einstein's
 work. It is interesting to note that Einstein predicted the so
 called Brownian motion of suspended microscopic particles not knowing
 that R. Brown first discovered it in 1827 [24]. 
 The resemblance of the Schr\"odinger equation to the diffusion
 equation had lead physicists (including Einstein and Schr\"odinger)
 to attempt to connect quantum mechanics with an underlying stochastic
 process, the Brownian
 process.     
\par Nelson [1], in 1966, constructed the Schr\"odinger
 equation from
an analysis of Brownian motion by identifying the forward and backward
average velocities of a Brownian particle with the real and imaginary
parts of a wave function. He pointed out that the basic process
involved is defined non-relativistically, and can be used if
relativistic effects can be ``safely'' neglected. The development
of a relativistically covariant formulation of Brownian motion could
 therefore
provide some insight into the structure of a relativistic quantum theory.
 \par Nelson pointed out that the formulation of his stochastic
mechanics in the context of general relativity is an important open
 question [1].  The Riemannian metric
spaces one can achieve, in principle, which arise due to nontrivial
correlations between fluctuations in spatial directions, could,  in the
 framework of a covariant theory of Brownian motion, lead to
spacetime pseudo-Riemannian metrics in the structure of diffusion and
Schr\"odinger equations.
Morato and Viola [25] have recently constructed a relativistic
 quantum equation for a free scalar field. They assumed the
 existence of a 3D (spatial) diffusion in a comoving frame, a
 non-inertial frame in which the average velocity field of the
 Brownian particle (current velocity) is zero. In this frame the
 location of the Brownian particle in space  experiences Brownian
 fluctuations parametrized by the proper time of the comoving
 observer. They interpreted possible negative
 0-component current velocities with what they called `rare events',
 which are time reversed Brownian processes (a peculiarity arising in
 the relativistic treatment). The equation they achieved this way is
 approximately the Klein-Gordon equation. It is important to note that
 in the inertial frame they do not obtain a normal diffusion. This is
due to 
the
 fact that their process is stochastic only
 in three degrees of freedom and therefore is not covariant. In this
paper
 we shall study a manifestly covariant form of Brownian motion.
 \ par In a previous work [27] we introduced a new approach
to the formulation of relativistic Brownian motion in 1+1
dimensions. The process we formulate is a straightforward
generalization of the standard one dimensional
 diffusion to 1+1 dimensions (where the
 actual random process is thought of as a `diffusion' in the time
 direction
 as well as in space), in an inertial frame. The equation achieved is
 an exact Klein-Gordon equation. It is a relativistic generalization
 of Nelson's Brownian process, the Newtonian diffusion. In this work
 we review the relativistic Brownian process in 1+1 dimensions [27]
 where the inclusion of both spacelike and timelike
 motion for the Brownian particle (event) is considered; if the
 timelike motion is considered as ``physical'' the ``unphysical''
 spacelike motion is represented (through analytic continuation) by
 imaginary quantities. We extend the treatment to 3+1 dimensions using
 appropriate weights for the imaginary representations. The complete 
formalism then can be used to
 construct relativistic general covariant diffusion and Schr\"odinger
 equations with pseudo-Riemannian metrics which follow from the
 existence of nontrivial correlations between the coordinate
 random variables.
\par Finally, we discuss the possible implications of the process we
 consider (i.e. a relativistic stochastic process with Markov property
 which preserves macroscopic Lorentz covariance) on the entangled
 system, where we claim that though fluctuations which exceed the
 velocity of light occur, the macroscopic
 behavior dictated by the resulting Fokker-Planck equation is local.

\bigskip
\noindent {\bf II. Equations of Motion and Self-Interaction}
\smallskip
\par The Stueckelberg-Schr\"odinger equation which governs the
evolution of a quantum state over the manifold of spacetime was
postulated by Stueckelberg [6] to be, for the free particle,
$$ i{\partial \psi_\tau \over \partial \tau} = {p^\mu p_\mu \over
2M}\psi_\tau
\eqno(2.1)$$
where, on functions of spacetime, $p_\mu$ is represented by
$-i \partial/\partial x^\mu \equiv -i\partial_\mu$. 
\par  Taking into account
 full $U(1)$ gauge invariance, corresponding to the requirement that
 the theory maintain its form under the replacement of $\psi$ by
 $e^{ie_0 \Lambda}\psi$, 
 the Stueckelberg-Schr\"odinger 
equation (including a compensation field for the
$\tau$-derivative of $\Lambda$) is [10]
$$ \bigl(i{\partial \over \partial \tau} + e_0 a_5(x,\tau) \bigr) \psi_\tau (x) = 
  {(p^\mu - e_0a^\mu(x,\tau))(p_\mu - e_0 a_\mu(x,\tau)) \over 2M
}\psi_\tau(x), \eqno(2.2)$$
where the gauge fields satisfy
$$ a_\alpha' = a_\alpha + \partial_\alpha \Lambda $$
under gauge transformations generated by 
$$\psi' = e^{ie_0\Lambda}\psi , $$
for $\Lambda$ a differentiable function of $\{x^\mu, \tau\} \equiv
x^\alpha$ ( $\alpha = 0,1,2,3,5$),which 
may depend on $\tau$, and $e_0$ is a coupling
constant which we shall see has the dimension $\ell^{-1}$.
The corresponding classical Hamiltonian then has the form
 $$ K=  {(p^\mu - e_0a^\mu(x,\tau))(p_\mu - e_0 a_\mu(x,\tau)) \over 2M
}- e_0 a_5(x,\tau) , \eqno(2.3)$$
in place of $(2.1)$.  Stueckelberg [6] did not take into account this
full gauge invariance requirement, working in the analog of what is
known in the nonrelativistic case as the Hamilton gauge (where the
gauge function $\Lambda$ is restricted to be independent of time).
The equations of motion for the field variables are given (for both the
 classical and quantum theories) by [10]
$$ \lambda \partial_\alpha f^{\beta \alpha}(x,\tau) = e_0
j^\beta(x,\tau), \eqno(2.4)$$
where $\alpha,\, \beta = 0,1,2,3,5$, the last corresponding to the
$\tau$ index, $j^5 \equiv \rho$ is he density of events in spacetime, 
  and $\lambda$, of dimension $\ell^{-1}$, 
 is a factor on the terms $f^{\alpha
\beta} f_{\alpha \beta}$ in the Lagrangian associated with $(2.2)$
(with, in addition, degrees of freedom of the fields) required by
 dimensionality. 
 The field strengths are 
$$ f^{\alpha \beta} = \partial^\alpha a^\beta - \partial^\beta
a^\alpha, \eqno(2.5)$$
and the current satisfies the conservation law [10]
$$ \partial_\alpha j^\alpha(x,\tau) = 0; \eqno(2.6)$$
integrating over $\tau$ on $(-\infty, \infty)$, and assuming that
$j^5(x,\tau)$ vanishes (pointwise) at $\vert \tau \vert \rightarrow \infty$, 
one finds that 
$$\partial_\mu J^\mu(x) = 0, $$
where (for some dimensionless $\eta$) [9,11] 
$$ J^\mu(x) = \eta \int_{-\infty}^\infty \, d\tau \, j^\mu(x,\tau).
\eqno(2.7)$$  
We identify this $J^\mu(x)$ with the Maxwell conserved current.
In ref. [28], for example, this expression occurs with
    $$ j^\mu (x,\tau) = {\dot x}^\mu(\tau) \delta^4(x-x(\tau)),
\eqno(2.8)$$
and $\tau$ is identified with the proper time of the particle
 (an identification which can be
made for the motion of a free particle). The conservation of the integrated current
then follows from the fact that 
$$\partial_\mu j^\mu = {\dot x}^\mu(\tau) 
\partial_\mu\delta^4(x-x(\tau))= -{d \over d\tau} \delta^4 (x-x(\tau)),$$
 a total derivative; we assume that the world line runs to infinity
 (at least in the
 time dimension) and therefore the $\delta$-function vanishes at the end
 points[6,28], in accordance with the discussion above.
 \par As for the Maxwell case, one can write the
current formally in five-dimensional form
$$ j^\alpha={\dot x}^\alpha\delta^4(x(\tau)-x). \eqno(2.9)$$ 
For $\alpha = 5$, the factor ${\dot x}^5$ is unity, and this component therefore
represents the event density in spacetime.
\par  Integrating the $\mu$-components of
Eq. $(2.4)$ over $\tau$ (assuming $f^{\mu 5} (x, \tau) \rightarrow 0$ 
(pointwise) for $\tau \rightarrow \pm \infty$), we obtain the Maxwell
 equations with the Maxwell charge $e= e_0/\eta$ and
 the Maxwell fields  given by 
$$ A^\mu(x) = \lambda \int_{-\infty}^\infty a^\mu(x,\tau) \,d\tau. \eqno(2.10)$$
 A Hamiltonian of the form $(2.3)$ without $\tau$ dependence of the
fields, and without the $a_5$ terms, as written by Stueckelberg [6], 
 can be recovered in the limit of the zero mode of the fields (with $a_5 =0$)
in a physical state for which this limit is a good approximation
{\it i.e.}, when the Fourier transform of the fields, defined by
$$ a^\mu(x,\tau) = \int \,ds {\hat a}^\mu (x,s)
e^{-is\tau}, \eqno(2.11)$$
 has support only in a small  neighborhood $\Delta s$ of $s=0$.  The vector
potential then takes on the form
 $  a^\mu(x,\tau) \sim \Delta s {\hat a}^\mu(x,0)
 =(\Delta s / 2\pi \lambda) A^\mu(x)$, and we
identify $e= (\Delta s/2\pi \lambda) e_0$. The zero mode therefore
emerges when the inverse correlation length of the field $\Delta s$ is
 sufficiently small, and then  $ \eta = 2\pi \lambda/\Delta s $. We
remark that in this limit, the fifth equation obtained from $(2.4)$
decouples.  The Lorentz force obtained from this Hamiltonian, using
the Hamilton equations, coincides with the usual Lorentz force, and,
as we have seen, the generalized Maxwell equation reduce to the usual
Maxwell equations.  The theory therefore contains the usual Maxwell
Lorentz theory in the limit of the zero mode; for this reason we have
called this generalized theory the ``pre-Maxwell'' theory.
\par If such a pre-Maxwell theory really underlies the
standard Maxwell theory, then there should be some physical mechanism
which restricts most observations in the laboratory to be close to the
zero mode. For example, in a metal there is a frequency, the plasma
frequency, below which there is no transmission of electromagnetic
waves. In this case, if the physical universe is imbedded in a medium
which does not allow high ``frequencies'' to pass, the
pre-Maxwell theory reduces to the Maxwell theory. Some study has been
carried out, for a quite different purpose (of achieving a form of
analog gravity), of the properties of the
generalized fields in a medium with general dielectric tensor [29]; we
discuss this study in detail in a later section.
Moreover, as we describe below [9] the
 high level of nonlinearity of
the theory of the electric charge in interaction with itself 
 may  generate an approximate
effective reduction to Maxwell-Lorentz theory, with high frequency
chaotic behavior providing the regularization achieved by models of
the type discussed by Rohrlich [30].
\par  We remark that
integration over $\tau$ does not bring the generalized Lorentz force
into the form of the standard Lorentz force, since it is nonlinear,
and a convolution remains.  If the resulting convolution is trivial,
i.e., in the zero mode, the two theories then coincide.  Hence, we
expect to see dynamical effects in the generalized theory which are
not present in the standard Maxwell-Lorentz theory.  In the following,
we describe results we have obtained in a study of the
self-interaction of a classical relativistic charged particle. 
\par  Writing the Hamilton equations
$${\dot x}^\mu = {dx^\mu \over d\tau} = {\partial K \over\partial
p_\mu};
\, \, \, \,  {\dot
p}^\mu = {dp^\mu \over d\tau} = -{\partial K \over dx_\mu} \eqno(2.12)$$ 
 for the Hamiltonian $(2.3)$,
we find the generalized Lorentz force [9,10]
$$ M {\ddot x}^\mu = e_0 f^\mu\,_\nu {\dot x}^\nu +
f^\mu\,_5. \eqno(2.13)$$
Multiplying this equation by ${\dot x}_\mu $, one obtains
$$ M{\dot x}_\mu {\ddot x}^\mu  = e_0{\dot x}_\mu f^\mu\,_5; \eqno(2.14)$$
this equation therefore does not necessarily lead to a trivial
relation between $ds$ and $d\tau$. The $f^\mu \,_5$ term has the
effect of moving
the particle off-shell (as, in the nonrelativistic case, the energy is
altered by the electric field).
\par Let us now define
$$\varepsilon = 1 + {\dot x}^\mu {\dot x}_\mu = 1 -{ds^2 \over
d\tau^2}, \eqno(2.15)$$
where $ds^2 = dt^2 -d{\bf x}^2$ is the square of the proper
time. Since
${\dot x}^\mu = (p^\mu - e_0 a^\mu)/M$, if we interpret $p^\mu -e_0
a^\mu)(p_\mu - e_0 a_\mu) = -m^2$, the gauge invariant particle mass [10], then
$$ \varepsilon = 1- {m^2 \over M^2} \eqno(2.17)$$
measures the deviation from ``mass shell'' (on mass shell, $ds^2 =
d\tau^2$).
\par We see that the $a_5$ field is strongly associated with the mass 
distribution; its source is the event density (mass density in
spacetime), i.e.(in generalized Lorentz gauge $\partial_\alpha
a^\alpha =0$), 
$$ -\partial_\alpha \partial^\alpha  a^5 = e j^5 \equiv \rho.$$
\par We carry out a  power series expansion of the Green's function in
the neighborhood of the parameter $\tau$ locating the particle on its
worldline, 
and compute the field strengths entering into the generalized Lorentz
force.  This results in the system of equations [9]
 (using a cutoff $\mu$,
which we estimate to be  of the order of $10^{-23}$ seconds [31]
to avoid explicit singularities)
 $$\eqalign{ M(\varepsilon, {\dot \varepsilon}) {\ddot x}^\mu &= -{1 \over 2}
 {M(\varepsilon,{\dot\varepsilon})\over {1-\varepsilon}}{\dot \varepsilon} {\dot x}^\mu
 + F(\varepsilon) e^2\bigl\{ {\mathop{x}^{...}}^\mu+ {1 \over
 1-\varepsilon}{\dot x}_\nu {\mathop{x}^{...}}^\nu {\dot x}^\mu
 \bigr\} \cr &+ e_0 {f_{ext}}^\mu\,_\nu {\dot x}^\nu +
 e_0\Bigl({{\dot x}^\mu {\dot x}_\nu \over
 1-\varepsilon}+\delta^\mu\,_\nu
 \Bigr){f_{ext}}^\nu_5.\cr} \eqno(2.18)$$
for the orbits in spacetime.  We moreover obtain an autonomous
equation for the off-shell deviation,
$$  \bigl(1+2{F_5\over
F_3}\bigr){\mathop{\varepsilon}^{...}}-A{\ddot
\varepsilon}+ B{\dot \varepsilon}^2 +
 C {\dot \varepsilon}-D+E{\dot
\varepsilon}^3+I{\dot \varepsilon}{\ddot \varepsilon}=0.
 \eqno(2.19)$$ 
The coefficients are defined as follows
$$\eqalign{ A &= {2 \over F_3}\bigl( {M \over 2e^2} + F_2 \bigr)
+ {2 M(\varepsilon, {\dot \varepsilon}) \over e^2 F(\varepsilon)}-
{4M(\varepsilon,{\dot \varepsilon})F_5\over
2e^2F(\varepsilon)},\cr B&= {2F'_3 \over F_3^2}\bigl(F_2-
{M\over 2e^2} \bigr) - {2F'_2 \over F_3} + {2 \over {1-\varepsilon}}
{1 \over F_3} \bigl( {M\over 2e^2} + F_2 \bigr)-{M(\varepsilon,
 {\dot\varepsilon})\over
e^2F(\varepsilon)} {1 \over 1-\varepsilon}, \cr C&= {4
M(\varepsilon, {\dot\varepsilon}) \over e^2 F(\varepsilon)}
 {1 \over F_3}\bigl( {M\over
2e^2} + F_2 \bigr) - {2  \over F_3^2}F_1 F'_3-{2F_1 \over
(1-\varepsilon)F_3} + {2 \over F_3}F'_1 ,\cr D &= {4
M(\varepsilon, {\dot\varepsilon}) \over e^2 F(\varepsilon)}{F_1 \over F_3} \cr
E&={2F_4\over (1-\varepsilon)F_3}-2\bigl({F_4F'_3\over
{F_3}^2}-{F'_4\over F_3}\bigr)\cr I &= 2 \bigl({F'_5\over
F_3}+2{F_4\over F_3}-{F_5F'_3\over{F_3}^2}\bigr)-
{2F_5 \over(1-\varepsilon)F_3}.\cr} \eqno(2.20)$$
\par We have retained the function $M(\varepsilon, {\dot \varepsilon})$ in 
some of the coefficients for convenience, although it contains a part
 linear in ${\dot\varepsilon}$, and would therefore modify the
 definitions of the coefficients
somewhat (the form of $(4.6)$ would
 remain) if we were to redefine them to make all the derivative of
 $\varepsilon$ terms explicit. 
\par The basic functions in terms of which the $F_i$ above are defined,
extracted directly from the Green's function, are 
$$\eqalign{ f_1(\varepsilon) &= {{3\over
2}\ln\bigl\vert{1+\sqrt{\varepsilon} \over
 1-\sqrt{\varepsilon}}\bigr\vert \over{(\varepsilon)}^{5 \over 2}}-{3
 \over \varepsilon^2(1-\varepsilon)}+{2 \over
 \varepsilon(1-\varepsilon)^2}\cr  f_2(\varepsilon) &=
 {{3\over 2}\ln\bigl\vert{1+\sqrt{\varepsilon} \over
 1-\sqrt{\varepsilon}}\bigr\vert \over{(\varepsilon)}^{5 \over
 2}}-{1\over \varepsilon^2}-{2 - \varepsilon  \over
 \varepsilon^2(1-\varepsilon)}\cr  f_3(\varepsilon) &= \ -
 {{1\over 2}\ln\bigl\vert{1+\sqrt{\varepsilon} \over
 1-\sqrt{\varepsilon}}\bigr\vert \over\varepsilon^{3 \over
 2}}+{1\over
 \varepsilon(1-\varepsilon)}. \cr} \eqno(2.21)$$
\par  For either sign of $\varepsilon$, when $\varepsilon \sim 0$,
 $$\eqalign{ f_1(\varepsilon) &\sim  {8 \over 5} +{24 \over
 7}\varepsilon+{16 \over 3}{\varepsilon}^2+O(\varepsilon^3), \cr
 f_2(\varepsilon) &\sim  -{2 \over 5} - {4 \over 7}\varepsilon-{2
 \over 3}{\varepsilon }^2+O(\varepsilon^3),\cr
 f_3(\varepsilon) &\sim  {2 \over 3} + {4 \over 5}\varepsilon+{6
 \over 7}{\varepsilon}^2+O(\varepsilon^3). \cr} \eqno(2.22)$$         
and therefore the functions defined in $(2.21)$ are smooth near
 $\varepsilon=0$.  We then define the auxiliary functions
 $$\eqalign{ g_1 &= f_1-f_2-3f_3 \,\, , \,\,g_2={1 \over
2}f_1-f_2-2f_3 \,\, , \,\,g_3={1 \over 6}f_1-{1 \over 2}f_2-{1 \over
2}f_3 \cr  h_1 &={1 \over 2}f'_1-{1 \over 2}f'_2-f'_3 \,\, ,
\,\,h_2={1 \over 4}f'_1-{1 \over 2}f'_2-{1 \over 2}f'_3\,\, ,
\,\,h_3=(f'_1-f'_2-f'_3) \cr
 h_4 &= f''_1-f''_2-f''_3.\cr}.\eqno(2.23)$$
  Finally,
$$\eqalign{ F_1(\varepsilon) &= { g_1(\varepsilon-1)\over
3\mu^2}\cr F_2(\varepsilon) &= {g_2-2(\varepsilon-1)h_1
\over 4\mu} \cr F_3(\varepsilon) &= g_3+{1 \over
12}(\varepsilon-1)h_3 \cr F_4(\varepsilon) &={1\over
2}h_2+{1\over 8}(\varepsilon-1)h_4 \cr F_5(\varepsilon)
&={1\over 8}(1-\varepsilon)h_3 \cr}\eqno(2.24)$$
and
$$ F(\varepsilon)={f_1 \over 3}(1-\varepsilon)+g_3.\eqno(2.25)$$ 
\par Here, the coefficients of ${\ddot x}^\mu$ have been grouped
formally into a renormalized (off-shell) mass term, defined (as done
in the standard
 radiation reaction problem ) as
$$ M(\varepsilon, {\dot \varepsilon}) = M + {e^2 \over
2\mu}\bigl[{f_1(1-\varepsilon) \over 2}+g_2\bigr]-e^2\bigl[{1\over
4}f_1'(1-\varepsilon)+h_2\bigr]{\dot
 \varepsilon}, \eqno(2.26)$$ 
where [9]
$$ e^2 =  {2e_0^2 \over \lambda (2\pi)^3\mu} \eqno(2.27)$$
 can be identified with the Maxwell charge by studying the on-shell
 limit (compare with our discussion above; one concludes that in this
case $\Delta s^2 \cong \lambda/2\pi\mu$). If $\varepsilon$ varies
slowly, this ``renormalized mass''
term can have some significance as a measurable mass, but if
$\varepsilon$ is rapidly varying, the identification is only formal;
its connection with a measurable mass is only in the sense of some
average over local variations.  
\par We remark that when one multiplies Eq.(2.18) by ${\dot x}_\mu$, it
 becomes an identity (all of the terms except for $ e_0 {f_{ext}}^\mu_
 \nu{\dot x}^\nu$ may be grouped to be proportional to $\Bigl( {{\dot
 x}^\mu{\dot x}_\nu \over 1-\varepsilon}+\delta^\mu_\nu \Bigr)$); one
 must use Eq.$(2.19)$ to compute the off-shell mass shift
 $\varepsilon$ corresponding to the longitudinal degree of freedom in
 the direction of the four velocity of the particle.  Eq.$(2.18)$
 determines the motion orthogonal to the four
 velocity. Equations $(2.18)$ and $(2.19)$ are the
 fundamental dynamical equations governing
 the off-shell orbit. 
\par  It can
be shown [9] that Eq.$(2.18)$ reduces to the ordinary
 (Abraham-Lorentz-Dirac)
 radiation
reaction formula for small, slowly changing $\varepsilon$ and that
that no instability, no radiation, and no acceleration of the
 electron occurs when it is precisely on shell. There is therefore no ``runaway
 solution'' for the exact mass shell limit of this theory.  This
 result indicates that in the mass shell
 limit, the theory is fundamentally different than the ``standard'' theory.
 The unstable Dirac result is approximate for $\varepsilon$ close to, but
 not precisely zero. The Dirac instabilities are therefore necessarily
 associated with at least small deviations from mass shell. 
\par The results of this calculation demonstrate the strong connection
 between the $a_5$ field, whose source is the mass distribution,  and
 the off-shell mass variations of the particle.  In the next section
 we give an explicit geometrical interpretation, constructing a
 conformal metric which replaces the fifth gauge field (as in standard
 electromagnetism, the scalar gauge field cannot be removed in the
 presence of sources). 
\bigskip
\noindent{\bf III. Conformal Equivalence for the Fifth Potential}
\smallskip
\par  In this section, we study the replacement of the fifth potential
 in a flat space picture in terms of a
new Hamiltonian which contains only a 4-potential, and takes into
account the fifth potential with a metric coefficient [11].
 The generator of motion is (we assume no explicit $\tau$ dependence
in the fields in this section, so that $a_\mu(x)$ is a zero-mode field)
$$ K_r=g^{\mu \nu}{(p_\mu-ea_\mu( x))(p_\nu-ea_\nu( x))
 \over 2M} \eqno(3.1)$$ 
\par The conformal structure of $g^{\mu\nu}$ follows from the fact that
in $(2.3)$, the contraction of ${(p_\mu-ea_\mu( x))(p_\nu-ea_\nu( x))
 \over 2M}$ with the Minkowski metric is just $K+ ea_5$, and hence the
 conformal factor $\Phi$ is given by he expression preceding $(3.5)$.
\par We now verify that this functional of $(p_{\sigma}, x^{\sigma})$
gives Hamilton equations which are equivalent dynamically to those of
$(2.3)$. We find
$$ { d x^{\sigma} \over d\tau}\equiv {\dot x}^\sigma= {\partial K_r \over \partial
p_\sigma}
=g^{\sigma \nu}{\dot \xi}_\nu, \eqno(3.2)$$
where we have defined the auxiliary variable $\xi$ by
    $$ {\dot \xi}_\nu = p_\nu -ea_\nu.$$
Note that the covariant Poisson bracket
$$ M \{{\dot \xi}_\nu, x^\lambda\}= \{p_\nu, x^\lambda\} =
\delta_\nu\,^\lambda,$$
so that 
 $$ \{{\dot x}^\sigma, x^\lambda\}= g^{\sigma\lambda} \{{\dot
\xi}_\nu, x^\lambda \}= {1 \over M}g^{\sigma\lambda}.$$
This type of commutation relation, a generalization [10] of the assumption
made originally by Feynman, maintains
 the interpretation of $p_\nu$ as the generator of translations in
$x^\lambda$ (in the corresponding quantum theory, $p_\nu \rightarrow
-i\partial/\partial x^\nu$.  We shall return to this point in a more
general context later.  
\par This equation gives us the transformation law $dx^\sigma=g^{\sigma
\nu}{d \xi_{\nu}}$ and  $d{\xi}_\sigma=g_{\sigma
\mu}d x^{\mu}$.  The second Hamilton equation gives
$${dp_\sigma \over d\tau}=-{\partial K_r \over \partial {
x^\sigma}}=-{M \over 2}{\partial g^{\mu \nu} \over \partial
x^\sigma} {\dot \xi}_{\mu}{\dot \xi}_{\nu}+g^{\mu \nu}{\partial
a_{\mu} \over \partial 
 x^\sigma}{(p_{\nu}-ea_{\nu})\over M} \eqno(3.3) $$
We now replace $  x$ with $\xi$ using the transformation law and
 substitute for ${\dot p}_{\sigma}$ to obtain
$$e{\dot \xi^{\mu}}{\partial a_{\mu} \over \partial
\xi^\sigma}+e{\partial a_5 \over \partial \xi^\sigma}=-{M \over
2}{\partial g^{\mu \nu} \over \partial {{\xi}_\alpha}}g_{\alpha
\sigma} {\dot \xi}_{\mu}{\dot \xi}_{\nu}+g^{\mu \nu}e{\partial a_{\mu}
\over \partial  {\xi}_\alpha}g_{\alpha \sigma}{\dot \xi}_{\nu}
 \eqno(3.4) $$
\par The functionals $K, K_r$ are different; however, on the physical
 trajectories
they take the same numerical value, $\bf{K}$.
We now show that choosing a conformal metric $g^{\mu \nu}={\Phi}(
x)\eta^{\mu \nu}$, the Lorentz force derived from $K_r$ is the same
as the one
 derived from $K$.
In this case we have
$$ g_{\mu \nu}={1 \over \Phi( x)}\eta_{\mu \nu}, \,\,\,\,\,\,\,
{\Phi}( x)={1 \over 1+{e\over \bf{K}} a_5( x)}. $$
and Eq.($2.8$) gives
$$ e{\partial a_5 \over \partial \xi^\sigma}=-{M \over 2}{1 \over
\Phi}{ \partial \Phi\over \partial \xi_{\sigma}} \eta^{\mu \nu}
 {\dot \xi}_{\mu}{\dot \xi}_{\nu}. \eqno(3.5) $$
Using ${M \over 2}\eta^{\mu \nu}{\dot \xi}_\mu{\dot \xi}_\nu={\bf{K}
\over \Phi}$ we find that Eq. $(2.9)$ is indeed satisfied. This shows
that the  Hamilton equations for the two generators are identical.
\par It is now interesting to examine the geodesic motion of this dynamical
system, assuming the fields are static in $\tau$. 
The Lagrangian in this case is 
$$L=p_{\mu}{\dot x}^\mu-H= {M \over 2}g_{\mu \nu}{\dot x}^\mu{\dot
x}^\nu+{\dot x}^\mu a_\mu. \eqno(3.6) $$
We now make a small variation in $x^\mu$
$$ x^\mu \rightarrow x^\mu+\delta x^\mu$$
$$ \delta S =\int d\tau \bigl[ {M \over 2} \bigl ({\partial g_{\mu \nu} \over \partial x^\sigma}{\dot x}^\mu{\dot x}^\nu \delta x^\sigma+2g_{\mu \nu}{\dot x}^\mu{d \delta x^\nu \over d\tau} \bigr)+ea_\mu{d \delta x^\mu \over d\tau}+{\partial a_\mu \over \partial x^\sigma}{\dot x}^\mu{\delta x}^\sigma \bigl]$$
From the minimal action principal we obtain, by  integration by parts
of the $\tau$ derivatives ($\tau$ independence of the field implies 
${d \over d\tau}={\dot x}^\mu {\partial \over \partial x^\mu}$)
$$ 0= {1 \over 2} \bigl ({\partial g_{\mu \nu} \over \partial
x^\sigma}{\dot x}^\mu{\dot x}^\nu-2g_{\sigma \nu}{\dot x}^\mu{\dot
x}^\sigma-2g_{\sigma \nu}{\ddot x}^\nu \bigr)+{e\over M}\bigl(-{\dot
x^\mu}{\partial a_\sigma\over
 \partial x^\mu}+{\partial a_\mu \over \partial x^\sigma}{\dot x}^\mu \bigl);$$
multiplying by $g^{\lambda \sigma}$ we finally get
$$ {\ddot x}^\lambda=-{\Gamma}^{\lambda}_{\mu \nu}{\dot x}^\mu{\dot
x}^\nu+{e \over M}{\dot x}^\mu f^{\lambda}_{\, \mu} \eqno(3.7)$$
where $f^{\lambda}_{\,\mu}=g^{\lambda \sigma}f_{\sigma \mu} $ and
$f_{\sigma \mu}={\partial a_\mu \over \partial x^\sigma}-{\partial
a_\sigma\over \partial x^\mu}$.
\par As an example of the application of our result for the conformal
metric, let us consider the Friedmann-Robertson-Walker universe.
\par In the ``flat space'' Robertson-Walker model (see,e.g., [32]), 
for the spatial geometry characterized by k=0, the metric
$$ds^2= d\tau^2-\Phi^2(\tau)\bigl(dx^2+dy^2+dz^2\bigr). \eqno(3.8)$$
 can be brought to the form
$$ds^2=\Phi^2(t)\bigl( dt^2-dx^2-dy^2-dz^2\bigr). \eqno(3.9)$$
 by using the transformation
$$t=\int{d\tau \over \Phi(\tau)}; \eqno(3.10)$$
$\tau$ is the time coordinate of a freely-falling object and therefore
coincides with our notion of universal the $\tau$. The function
$\Phi(\tau)$ is often designated by $R$ or $a$ and is the
(dimensionless) spatial scale of the expanding universe.
In the conformal coordinates the time-coordinate is therefore related 
to $\tau$, according to the transformation above, through
$${dt \over d\tau}= {1 \over \Phi}. \eqno(3.11)$$
 It is interesting to use the Lorentz force in order to achieve the
same result. Let us assume that $a_5$ depends on $t$ alone.  In this
case, the force is
$$ {\ddot t}={e \over M}f^0_{\,5}=-{e \over M}{da^5 \over dt} \eqno(3.12)$$
The relation
$$ \Phi^2={1 \over 1+{e \over \bf{K}}a^5} \eqno(3.13)$$
then implies
$$ 2 {d\Phi \over dt}\Phi=-\Phi^4{e \over \bf{K}}{da^5 \over dt},$$
i.e.,
$${da^5 \over dt}={ {\bf K}\over e}{d \over dt}\bigl({1 \over \Phi^2}\bigr)$$
We substitute this in the force equation and multiply by $2 {\dot t}$
to obtain
$$ {d {\dot t}^2\over d\tau}=-2{{\bf K}\over M}{d \over d\tau}\bigl({1
\over \Phi^2}\bigr). \eqno(3.14)$$
Finally,  putting $\bf{K}=-{M \over 2}$ we arrive at the remarkable result
$${dt \over d\tau}={1 \over \Phi},$$
which coincides with the transformation $(3.11)$ from the time on the
freely falling clock $\tau$ to the redshifted $t$ in the conformal
form of the Robertson-Walker metric.  We see that this $t$ corresponds
to the Einstein time satisfying the dynamical Hamilton equations, and
the conformal factor of the Robertson-Walker metric coincides with the
conformal facter of the curved space embedding.
\par In this construction we have assumed the $a_5$ field to depend on
$t$ alone.  The generalized Maxwell equations then provide a simple
connection between the Robertson-Walker scale and the event
density.
\par The generalized Maxwell equations [10] are
$$ \partial_\alpha f^{\beta \alpha} = ej^\beta, \eqno(3.15)$$
where $j^\beta = (j^\mu, \rho)$ satisfies $\partial_\beta j^\beta =
\partial_\mu j^\mu + \partial_5 \rho = 0$, and $\rho$ is the event
density.  In the generalized Lorentz gauge $\partial_\alpha a ^\alpha
=0$, we have
$$ -\partial_\alpha \partial^\alpha a^5 = ej^5 = e\rho. \eqno(3.16)$$
Since $a^5$ depends on t alone $(3.16)$ becomes
$$ \partial_t^2  a^5 = e\rho. \eqno(3.17)$$
From $(3.13)$, 
$$ a_5={{\bf K} \over e}\bigl({1 \over \Phi^2}-1\bigr)$$
so that from $(3.17)$
$$\rho=-{2 {\bf K} \over e^2} \bigl[ {\Phi_{tt} \over
\Phi^3}-3{\Phi_t^2
 \over \Phi^4} \bigr] \eqno (3.18)$$
\par  The space-time geometry is related to the density of matter
$\rho_M$ through the Einstein equations
$$ G^{\mu \nu}\equiv R^{\mu \nu}-{1 \over 2}g^{\mu \nu}R=8 \pi G
T^{\mu \nu}, \eqno(3.19) $$ 
where $R^{\mu \nu}$ is the Ricci tensor, $R$ is the scalar curvature
and $T^{\mu \nu}$ is the energy-momentum tensor. For the perfect fluid
model (isotropy implies the $T^{\mu \nu}$ is diagonal)
$$T^{\mu \nu}=\rho u^\mu u^\nu+P(g^{\mu \nu}+u^\mu u^\nu). \eqno(3.20)$$
The $(0,0)$ component (referring to $\tau$) is then 
$$T^{\tau \tau}=8 \pi G \rho_M. \eqno(3.21)$$
using the affine connection derived from the metric $(3.8)$ one finds
$$ G^{\tau\tau}= 3 {{\dot{\Phi}}^2 \over \Phi^2}=8 \pi G \rho_M \eqno(3.22)$$
and the (equal) diagonal space-space components are (for example, we
write the $x,x$ component
$$G^{xx} = - {1 \over \Phi^2}\bigl[2 {{\ddot a}\over a}+{{\dot a}^2
\over a^2}\bigr]=8 \pi G T^{xx} = {8 \pi G P \over
\Phi^2}. \eqno(3.23)$$
\par Since $T^{\tau\tau}$ is the $(0,0)$ component of a tensor, it
follows from $(3.11)$ and $(3.21)$ that the matter density in the conformal
coordinates is given by
$$ \rho'_M = { 1 \over \Phi^2} \rho_M. \eqno(3.24)$$
To establish a connection between the density of events in spacetime
$\rho$ and the density of matter(particles) $\rho'_M$, in space at a
given time $t$, we assume that,
$$ \rho'_M=\rho \Delta t \eqno(3.25)$$
where $\Delta t$ is the time interval (in the conformal coordinates associated with the Stueckelberg evolution) in which the events generating the particle world lines are uniformly spread. 
It then follows from $(3.25)$ that 
$$  \Delta t = {\rho_M \over \rho \Phi^2}. \eqno(3.26)$$
\par We now consider two examples.  For the static universe, for $\rho_M$
constant, it follows from Eq. $(3.22)$ that $\Phi$ is given by an
exponential; it then follows from $(3.18)$ that $\rho$ is constant, so
that 
$$ \Delta t \propto \Phi^{-2}. \eqno(3.27)$$
\par For the matter dominated universe, where the pressure is
negligible [32], one sees from $(3.23)$ that 
$$ 2\Phi \Phi_{tt}= \Phi_t^2, $$
and substituting in $(3.18)$, one finds after changing $\tau$
derivatives to $t$ derivatives in $(3.22)$ that ${\rho_M \over \rho}$ is
constant.  It then follows that $\Delta t \propto \Phi^{-2}$ in this
case as well. 
\par This result implies that, at any given stage of development of
the universe, i.e., for a given $\tau$, the events generating the
world lines lie in an interval of the conformal time $t$ which becomes
smaller as $\Phi$ becomes large in the order of $\Phi^{-2}$ With the
relation $(3.11)$, this corresponds, on the other hand, to a narrowing
distribution, of order $\Phi^{-1}$ in $\tau$, contributing to a set of
events observed at a given value of the conformal time
$t$.   In general, if one observes the configuration of a system at a
given $t$, the events detected may have their origin at widely
different values of the world time $\tau$ parametrizing the
trajectories (world lines) of the spacetime events.  It would be
generally difficult to relate such configurations to the
configurations in spacetime (at a given $\tau$, instead of at a given
$t$) predicted by a dynamical theory.  However, in this case, we see
that the spreading is narrowed for large $\Phi$, so that the set of
events occurring at a given $\tau$ is essentially the same as the set
of particles occurring at a given $t$.  The observed configuratoins
therefore become very close to those predicted by the underlying
dynamical model. In the general case, the relation between $\rho
(\tau)$ and $\rho_M(t)$ could be very complicated, and it may be
difficult to see in the observed configurations a simple relation to
the dynamical model evolving according to the world time.
 In the static and matter dominated Friedmann-Robertson-Walker model, the
correspondence between the dynamical theory and observed
configurations becomes more clear as $\Phi$ becomes large.
 \par We have shown that the fifth potential of the generalized Maxwell
theory, obtained throught the requirement of gauge invariance of the
Stueckelberg-Schr\"odinger equation, can be eliminated in the function
generating evolution of the classical system by replacing the
Minkowski metric in the kinetic term by a conformal metric.  The
Hamilton equations resulting from this function coincide with the
geodesic associated with this metric, and with the Hamilton equations
of the original form, i.e., the geodesic equations of the conformal
metric describe orbits that coincide with solutions of the original
Hamilton equations, as found in previous work which studied the
replacement of an invariant (action-at-a-distance) potential by a
conformal metric [33].  In this case the geodesic equations are those
obtained from the conformal geodesic with the addition of a Lorentz
force in standard form.
\par The Robertson-Walker metric can be put into conformal form. The
conformal factor of the Robertson-Walker metric can then be put into
correspondence with the $a^5$ field of the generalized Maxwell theory
and therefore, through its $t$ derivatives (we assume no explicit
$\tau$ dependence) with the event density. In both the static and the
matter dominated models, the set of events generating the world
lines of the expanding universe condense into progressively thinner
slices of the conformal time.
\par We now consider a framework in which one can construct an analog  model
for gravity of much greater generality. 
\bigskip
\noindent{\bf IV. Eikonal Approximation to Wave Equations}
\smallskip
\par In this section, we derive the relation between the eikonal
 approximation  to the
five dimensional generalization of Maxwell theory required by the
gauge invariance of Stueckelberg's covariant classical and quantum
dynamics to demonstrate, in this approximation, the existence
of geodesic motion for the flow
of mass in a four dimensional pseudo-Riemannian 
manifold.  These results provide a
 foundation for the 
geometrical optics of the five dimensional radiation theory and
establish a model in which there is mass flow along
geodesics. 
\par We  apply this method as well to  the interesting case of 
relativistic
quantum theory with a general second rank tensor coefficient, as in
the conformal case studied above.  In this case the
eikonal approximation to the relativistic quantum mechanical current
 coincides with  the 
geodesic flow governed by the pseudo-Riemannian metric obtained from the
 eikonal approximation to solutions of the Stueckelberg-Schr\"odinger
equation.
 This construction provides a model in which there is an
underlying quantum mechanical structure for classical dynamical
motion along geodesics on a pseudo-Riemannian manifold. The
locally symplectic structure which emerges is that of Stueckelberg's
covariant mechanics on this manifold, and provides the principal
result of this work.
  We apply a technique similar to that used by Kline and Kay [13] in  the
study of four dimensional (Maxwell) wave equations to study
 the structure of wave equations in five dimensions [34] which
follow as a consequence of gauge invariance of the covariant classical
and quantum mechanics of Stueckelberg [6]. Since the
eikonal approximation results in the loss of one dimension (a high
frequency limit), the eikonal method applied to four dimensional wave
equations in a medium with non-trivial dielectric tensor results in a
ray approximation on a manifold in three dimensions (Riemannian).  In
the case of the five dimensional radiation theory, one finds a ray
approximation corresponding to geodesic flow in a four dimensional
pseudo-Riemannian manifold. We show that
there is a Hamiltonian form for the generation of the rays.
\par  We shall use, in this section,
Greek letters for  space time indices
 ($\mu=0,1,2,3$) and Latin letters to include
a fifth index representing the Poincar\'e invariant $\tau$ parameter in
addition to the usual 4 spacetime coordinates (e.g., $q=0,1,2,3,5$). 
The analysis proceeds by a generalization of the method [29] of replacing
 the electric and magnetic vector fields by the
electromagnetic tensor fields $(E,B)$ and the excitation tensor fields
 $(D,H)$. 
The generalized electromagnetic field tensor is  written
$$f_{q_1\, q_2} \equiv {\partial}_{q_1}a_{q_2}-{\partial}_{q_2}a_{q_1},  $$
where $a_q$ are the so-called pre-Maxwell electromagnetic
potentials (as pointed out above, the fifth gauge potential $a_5$ is
required for gauge
compensation of $i\partial_5$, generating the  evolution of the Stueckelberg
wave function [10]).
\par We introduce the dual (third rank) tensor
$$ k^{l_1 \, l_2 \, l_3}=\varepsilon^{\, l_1 \, l_2 \, l_3 \, q_1 \,
q_2}f_{\, q_1 \, q_2},$$
where $\varepsilon^{\, l_1 \, l_2 \, l_3 \, q_1 \,
q_2}$ is the antisymmetric fifth rank Levi-Civita tensor density.
The homogeneous pre-Maxwell equations are then given by
$${\partial}_{l_3}k^{\, l_1 \, l_2 \, l_3}=0, \eqno(4.1)$$
or, more explicitly ($\partial_5 = \pm \partial/\partial\tau$, according to
the signature of the $\tau$ variable, i.e., corresponding, as we have
discussed above, to $O(4,1)$
or $O(3,2)$ symmetry of the homogeneous field equations),
$${\partial}_5 \varepsilon^{l_1 \, l_2 \, 5 \, q_1 \, q_2}f_{q_1 \, q_2}+{\partial}_{\sigma} \varepsilon^{l_1 \, l_2 \,\sigma \, q_1 \, q_2 }f_{q_1 \, q_2}=0 . \eqno(4.2)$$ 
We now divide Eq.$(4.2)$ into two cases. In the first, the indices $l_1, l_2$
 correspond only to space-time indices: 
$$ {\partial}_{5} \varepsilon^{\mu \nu \, 5 \lambda \sigma
}f_{\lambda \sigma }+2{\partial}_{\sigma} \varepsilon^{\mu \nu \sigma
\lambda 5}f_{\lambda 5}=0 \rightarrow \,{\partial}_{5}
\varepsilon^{\mu \nu \,  \lambda \sigma }f_{\lambda \sigma
}+2{\partial}_{\sigma} \varepsilon^{\mu \nu \sigma \lambda }f_{\lambda
5}=0, \eqno(4.3) $$
where $\varepsilon^{\mu\nu\sigma\lambda}$ is the four dimensional
Levi-Civita tensor density
This equation, on the 0-mode ($\tau$ independent Fourier components)
does not involve any of the usual Maxwell fields but only the fifth
(Lorentz scalar) electromagnetic field. The second set from Eq.$(4.2)$
corresponds to $l_1 \, {\rm or}\,  l_2=5 $. It is clear then that all the
 other 4-remaining indices must be space-time indices and we obtain
$$ \varepsilon^{5 \mu \sigma \delta \nu}{\partial}_{\sigma} f_{\delta
\nu}=0
 \rightarrow \,  \varepsilon^{\mu \sigma \delta
\nu}{\partial}_{\sigma} f_{\delta \nu}=0 . \eqno (4.4)$$ 
It is this equation that reduces on integration over all $\tau$, to
the two usual homogeneous Maxwell equations.  The operation of
integrating over all $\tau$ extracts the zero frequency components (in
$\tau$), which we call the $0$-mode. This has the effect of reducing
the pre-Maxwell system of equations, as we discuss below, to the usual
 Maxwell equations. With appropriate identification of the
integrated quantities, the zero mode of the pre-Maxwell equations
coincides with the Maxwell theory (it is for this reason that the five
dimensional gauge fields associated with the Stueckelberg theory are
called ``pre-Maxwell'' fields).
\par We now turn to the current dependent pre-Maxwell equations. These
can be written as
$${\partial}_{l_1}n^{l_2 \, l_1}=-j^{l_2}, \eqno(4.5)$$
where $n^{l_1 l_2}$ are the matter induced (excitation) fields
 (corresponding to ${\bf H}, {\bf D}$ in the 4D theory). 
We remark that, restricting our attention to the spacetime components
of Eq. $(4.5)$, which then reads
$$ \partial_5 n^{\mu 5} + \partial_\nu n^{\mu \nu} = -j^\mu, \eqno(4.6)$$
we may extract the 0-mode by integrating over all $\tau$.  Since
$j^k$ satisfies the five dimensional conservation law $\partial_k j^k
=0$, its integral over $\tau$ (assuming$^{12}$  $j^5 \rightarrow 0$ for $\tau
\rightarrow \pm \infty$) reduces to the four dimensional conservation
law $\partial_\mu J^\mu =0$, where $J^\mu = \int j^\mu(x,\tau)d\tau$
is
 the 0-mode part of
$j^\mu$ (this formula for the conserved $J^\mu$ is given in
Jackson [28]). 
 The first term of the left side of $(4.6)$ vanishes 
(assuming that $n^{\mu 5}\rightarrow 0$ for $\tau \rightarrow \pm
\infty$), and one obtains the form
$$ \partial^\nu F^{\mu \nu} = J^\mu,$$
where we may identify the zero mode fields $F^{\mu \nu}$, as above, with the
Maxwell fields ${\bf H}, {\bf D}$.  With the zero mode of $(4.4)$, we
see that the Maxwell theory is properly contained in the five
dimensional generalization we are studying here.
\par We assume 
the existence of linear constitutive equations in the dynamical
structure of the 5D fields  in a medium 
which connects the $n$ tensor-field to the $k$
tensor-fields using a fifth rank tensor ${\cal E}$ which is a
generalization of the fourth rank covariant permeabilty- dielectric
tensor [29] which relates the $E,B$ fields to the excitation fields $D,H$
in the usual Maxwell electrodynamics. The constitutive equations have the form
$$n^{l_1 \, l_2}={\cal E}^{l_1 \, l_2 \, q
_1 \, q_2 \, q_3}k_{q_1 \, q_2 \, q_3}, \eqno (4.7)$$
antisymmetric in $l_1 l_2$ as well as $q_1 q_2 q_3$ (the indices of
$k$ have been lowered with the Minkowski metric tensor; we shall treat
other tensors in the same way in the following).
It is useful at this point to  distinguish between the space-time
elements 
 $f_{\mu \nu}$ and the elements $f_{\mu 5}$, and assume that the
tensor introduced in Eq. $(4.7)$ does not mix these fields (for
$n^{\mu5}$, if $\varepsilon_{\mu 5 q_1 q_2 q_3}$ has $q_1 q_2 q_3 =
\alpha \beta \gamma$, then the components of $k$ that enter are 
of the form $k_{\alpha \beta \gamma} = {\cal E}_{\alpha \beta \gamma
\mu 5} f^{\mu 5}$ only; similarly, for $n_{\mu \nu}$, only the
components ${\cal E}_{\mu \nu \alpha \beta 5}$ can occur, and
$k^{\alpha \beta 5}$ connects only to the components $f_{\lambda
\sigma}$ of the field tensor). As we have pointed out above, the
vector $f_{\mu 5}$ is physically distinguished from the antisymmetric
tensor $f_{\mu\nu}$ in the inhomogensous field equations, since the
source terms break the higher symmetry of the homogeneous field
equations. The assumption that the constitutive equations do not
couple these components results in a simpler system to analyze,
although (as for Hall type effects in the non-relativistic theory) it is
conceivable that the more general case could occur.
\par We introduce the
new set of fields:
$$b^{\mu \nu}={1 \over 2}\varepsilon^{\mu \nu \lambda
 \sigma}f_{\lambda \sigma},\eqno(4.8)$$
so that 
$$ n^{\lambda \sigma} = 2 {\cal E}^{\lambda \sigma \alpha \beta
5}b_{\alpha \beta}.\eqno(4.9)$$
On the zero mode, the fields $b_{\mu \nu}$ correspond to the dual
Maxwell fields; in
this theory they play a role analogous to the ${\bf B}$ fields in the
Maxwell theory. In a similar way, the $f_{\mu5}$ fields are analogous
to ${\bf E}$. The part of the tensor ${\cal E}_{l_1 \, l_2 \, q
_1 \, q_2 \, q_3}$ connecting the $\mu 5$ fields is discussed below.
\par Working with these fields enables us to construct the
equations in a form which, as we shall show, generalizes the Maxwell
theory to a form where the invariant time $\tau$ plays the role of $t$
and spacetime plays the role of space. This analogy helps to interpret
the physics and it distinguishes between the familiar physical
quantites $f_{\mu \nu}$ and the new fields $f_{\mu 5}$.    
Substituting these fields in $(4.3)$ and $(4.4)$,
we find 
$${\partial}_5 b^{\mu \nu }+{\partial}_{\sigma} \varepsilon^{\mu \nu
\sigma \lambda }f_{\lambda 5}=0, \eqno (4.10) $$
 and
$$ {\partial}_{\sigma} b^{\mu \sigma}=0. \eqno (4.11)$$
 For the spacetime excitation fields we define
$$h_{\mu \nu}={1 \over 2}\varepsilon_{\mu \nu \lambda
 \sigma}n^{\lambda \sigma}$$
and we get from $(4.5)$, for $l_2=\mu$,
$${\partial}_5n^{\mu 5}-{1 \over 2} \varepsilon^{\mu \sigma \lambda
\nu} {\partial}_\sigma h_{\lambda \nu}=-j^\mu, \eqno (4.12)$$
where we have used
$${\varepsilon_{\alpha \beta \eta \delta }}{ \varepsilon^{\eta \delta
\gamma \mu
}}=-2\bigl({\delta}_\alpha^\gamma{\delta}_\beta^\mu-{\delta}_\alpha^\mu{\delta}_\beta^\gamma
\bigr). \eqno(4.13)$$ 
To complete the set of equations, we note that for $l_2=5$, we get from $(4.5)$
$$\partial _\sigma n^{5 \sigma}=-j^5 .\eqno(4.14)$$
\par 
 To obtain a mass-energy conservation law for the fields, we multiply
 $(4.12)$ by $f_{\mu 5}$ and $(4.10)$ by $h_{\mu \nu}$, and then combine them, obtaining
$$ \bigl[f_{\mu 5} \partial_5 n^{\mu 5}+{1\over 2}h_{\mu \nu}{\partial}_\tau b^{\mu\nu}\bigr]+{1 \over 2} \varepsilon^{\sigma \mu \lambda \nu}{\partial}_\sigma (f_{\mu 5}h_{\lambda \nu})=-j^\mu f_{\mu 5}. \eqno (4.15)$$
Assuming the dielectric tensor reduced into the $\mu 5$ and $\mu \nu$
subspaces is symmetric (the relations of $n^{\mu 5}$ to $f_{\mu 5}$
and $n^{\mu \nu}$ to $f_{\mu \nu}$ go by the contraction ${\cal
E}\varepsilon$; the exclusive property of indices of $\varepsilon$ then
imply simple conditions on ${\cal E}$ for the symmetry of these forms) we can write $(4.15)$ as
$${1\over 2}{\partial}_5 \bigl[f_{\mu 5} n^{\mu 5}+{1\over 2}h_{\mu \nu} b^{\mu\nu}\bigr]+ {1 \over 2} \varepsilon^{\sigma \mu \lambda \nu}{\partial}_\sigma (f_{\mu 5}h_{\lambda \nu})=-j^\mu f_{\mu 5}.\eqno (4.16)$$
From the Stueckelberg Hamiltonian [10]
$$ K = {1 \over 2M} (p^\mu - e_0a^\mu)(p_\mu - e_0a_\mu) - e_0 a_5, \eqno(4.17)$$
it follows from the relativistic Lorentz force [9,10] 
$$M{\ddot x}^\mu = e_0 f^\mu\,_\nu {\dot x}^\nu + e_0 f^{\mu 5}, \eqno(4.18)$$
 as discussed above, that
$$ M{\dot x}_\mu{\ddot x}^\mu = M {d \over d\tau}( {\dot x}_\mu{\dot
x}^\mu) = {\dot x}_\mu f^{\mu 5} \eqno(4.19)$$
\par Since, in the Stueckelberg theory,
the Hamilton equations imply that 
$$ {\dot x}^\mu = {1 \over M}(p^\mu - e_0a^\mu),$$
and hence the $\tau$ derivative in the central equality of $(4.19)$
corresponds to the change in the mass-squared $(p^\mu -
e_0a^\mu)(p_\mu - e_0a_\mu)$ of the particle. 
It therefore follows that
 $j^\mu f_{\mu 5}$, is the rate of mass change of the
system.
 We
therefore  identify $s^\sigma=
{1 \over 2}\varepsilon^{\sigma \mu \lambda\nu}f_{\mu 5}h_{\lambda
\nu}$ as the analogue of the Maxwell Poynting vector. This
Poynting 4-vector is the mass radiation of the field.  We see,
furthermore, that ${1 \over 2}\bigl[f_{\mu 5} n^{\mu 5}+{1\over 2}h_{\mu \nu}
b^{\mu\nu}\bigr]$ is the scalar mass density of the field (its four
integral is the dynamical generator of evolution of the
non-interacting field [10]).
\par  We now introduce the eikonal approximation, i.e., set 
$$f_{l_1
l_2}(x,\tau)=f_{l_1 l_2}(x)\exp\,i\kappa( \tau-\Psi(x))$$
 for large $\kappa$. 
In the absence of
sources the 5D-Maxwell equations $(4.10), (4.11), (4.12), (4.14)$ take the form
(for large $\kappa$)
$$ b^{\mu \nu }-\varepsilon^{\mu \nu \sigma \lambda
}p_{\sigma}f_{\lambda 5}=0, \eqno (4.20) $$
$$ p_{\sigma} b^{\mu \sigma}=0 , \eqno (4.21)$$
$$n^{\mu 5}+{1 \over 2} \varepsilon^{\mu \sigma \lambda \nu} p_\sigma h_{\lambda \nu}=0, \eqno (4.22)$$ 
$$p_\sigma n^{5 \sigma}=0, \eqno(4.23)$$
where $p_\sigma=\partial_\sigma \Psi$.
\par
We now relate the direction of $p_\mu$ to the polarization of the
 fields. We write the ``cross product'' of $n$ and $b$ (analogous to
 the cross product
 of ${\bf D}$ and ${\bf B}$ in Maxwell's theory):
$$\varepsilon_{\mu \nu \sigma \lambda}n^{\nu 5} b^{\sigma \lambda}=2
n^{\nu 5}f_{\nu 5}p_\mu,\eqno(4.24)$$
or
$$p_\mu={1\over 2 n^{\alpha 5}f_{\alpha 5}}\varepsilon_{\mu \nu \sigma
\lambda}n^{\nu 5} b^{\sigma \lambda},$$
where we have used $(4.20)$ and $(4.23)$.
It is clear that since $p_\mu$ and the Poynting four-vector are cross
products of tensors which are not necessarily aligned in the same
four-directions, they are in general not parallel to each other (in
space-time) due to the anisotropy of the medium, i.e., the wave normal
and radiation flow directions are not, in general, the same.
\par The relations $(4.20)-(4.23)$, along with the constitutive relations
relating $n^{\mu \nu}, f_{\sigma \lambda}$, and $n^{\mu 5}, f_{\sigma
5}$, provide relations analogous to those of the standard Maxwell
theory characterizing the possible
 field strengths of the eikonal
approximation in terms of properties of the medium.  We shall not
treat these relations here, but discuss the mass-radiation
flows, along the rays,  on spacetime
 geodesics in the interesting special case where $h^{\mu \nu}=b^{\mu \nu}$,
which is analogous to the case of materials with $\mu=1$ in Maxwell's
electromagnetism. This case is interesting since, although the space is
empty in the usual sense(i.e. ${\bf E}={\bf D} \, ,\, {\bf B}= {\bf
H}$), the  dielectric effect involving the $f_{\mu 5} $ components can
drive the radiation on curved trajectories, i.e., the corresponding
spacetime can have a non-trivial metric structure.  
\par  We multiply $(37)$ by ${1 \over 2}\varepsilon^{\alpha \beta \mu
\nu}p_{\beta}$. We then use $(39)$ and $(30)$ to obtain
$$ {n_\alpha}^5-p_\alpha p^\beta f_{\beta 5}+p_\beta p^\beta
 f_{\alpha 5}=0 \eqno (4.25)$$
\par Defining the reduced dielectric tensor ${\cal E_\alpha}^\beta$ as the
part of the general dielectric tensor which connects only the $\alpha 5$ components of
the fields, i.e.,
$$n_{\alpha 5}={{\cal E}_\alpha}^\beta f_{\beta 5}, \eqno(4.26)$$
the condition $(40)$ then implies that ${\cal E}_\alpha^\beta f_{\beta
5}$ cannot be in the direction of $p^\alpha$ (unless it is lightlike).
We  obtain from Eq. $(42)$
$$\bigl({\cal E}_\alpha\,^\beta-p_\sigma p^\sigma {\delta_\alpha}^\beta +p_\alpha p^\beta \bigr)f_{\beta 5}=0, \eqno(4.27)$$
where we have chosen the negative sign for the signature of the fifth
 index, $n_{\alpha 5}=-{n_\alpha}^5$ [with this choice the flat space
limit, for which ${\cal E}_\alpha\,^\beta = \delta_\alpha\ ^\beta$,
Eq. $(4.27)$, with $(4.23)$, admits only spacelike $p_\alpha$; for
positive signature of the fifth index, in this limit, $p_\alpha$ would
be timelike].
\par Eq. $(4.24)$ has a solution only if the determinant of the coefficients
vanishes (a similar calculation in which the field strengths $f_{\mu
\nu}$ enter in place of $f_{\mu 5}$ results in the same condition on
these coefficients, as it must). It is somewhat simpler to work with
the eigenvalue equation
$(4.27)$. Assuming as before that this dielectric tensor is symmetric,
we can work
in a Lorentz frame in which it is diagonal.  In this frame we have
(for the transformed fields)
$$ f_{\alpha 5}= -{p_\alpha \over ({\cal E}^\alpha - p^2)}(p^\beta f_{\beta
5}). \eqno(4.28)$$
Note that in the isotropic case for which all of the ${\cal E}^\alpha$
are equal, one obtains $p_\beta f^{\beta 5}=0$, and
the metric becomes conformal, i.e., one obtains the condition 
$${\cal E}^{-1}\eta^{\mu\nu}p_\mu p_\nu = -1, $$
where $\eta^{\mu\nu}$ is the flat space Minkowski metric $(-1,1,1,1)$.
\par  Multiplying the equation $(4.28)$
on both sides by $p^\alpha$, and summing over $\alpha$, one obtains
the condition ($p^2 \equiv p_\mu p^\mu$),
$$ 0=K={p_1^2 \over {\cal E}_1-p^2} + {p_2^2 \over {\cal E}_2-p^2} +
{p_3^2 \over {\cal E}_3-p^2} - {p_0^2 \over {\cal E}_0-p^2} +1. \eqno (4.29)$$
 This condition determines, in this case, the Fresnel surface of the
wave fronts.
\par It then follows that
$$ {\partial K \over \partial p_\mu}={2 p^\mu \over {\cal E}^\mu -p^2}
+2 p^\mu {\partial K \over \partial p^2}. \eqno(4.30)$$  
 Calculating the  scalar product of $(4.28)$ and $(4.30)$ one then obtains
$$\eqalign{f_{\mu 5} &{\partial K \over \partial p_\mu} =\cr
&= -2(p_\nu
f^{\nu 5}){ \{ \sum_{i=1,2,3} {(p^i)^2 \over ({\cal E}^i-p^2)^2}-{{p_0}^2
 \over ({\cal E}^0-p^2)^2} -{\partial K \over \partial p^2}
 \}} = 0  \cr}. \eqno(4.31) $$
Multiplying the expression $(4.20)$ for  $b_{\mu \nu} (h_{\mu \nu})$ by
$(4.30)$, the contribution of the second term of $(4.30)$
vanishes since the Levi-Civita tensor is antisymmetric; the first
term, according to $(4.28)$, is proportional to $f^{\alpha 5}$, and
vanishes for the same reason. 
It therefore follows that
$$ {\partial K \over \partial p^\mu}h^{\mu \nu}=0. \eqno(4.32)$$
Since the scalar product of ${\partial K \over \partial p^\mu}$ with
both $h^{\mu\nu}$ and $f^{\mu 5}$ is zero, it is proportional to their ``cross
product'' i.e., it is parallel to the Poynting vector. To make the
proof explicit, it is convenient to define $V^\mu = {\partial K \over
 \partial p_\mu}$, $H_i = -h_{0j}$, $F_i = f_{i5}$, and $D^i =
\varepsilon ^{ijk}h_{jk}$ (the space index may be raised or lowered
without changing sign in our Minkowski metric).  In this case, the
conditions $V^\mu h_{\mu \nu} =0$ and $V^\mu f_{\mu 5}$ become
$$ \eqalign{V^0{\bf H} - {\bf V} \times {\bf D}&=0 \cr
-V^0 f^{05} + {\bf V}\cdot {\bf f} &=0 \cr
{\bf V} \cdot {\bf H} &=0, \cr} \eqno(4.33)$$
where we have used boldface to represent the space components of the vector.
In these terms, the Poynting vector is given by
$$ \eqalign{S^0 &=  {\bf D}\cdot {\bf f}\cr
{\bf S} &= f^{05} {\bf D} + {\bf f} \times {\bf H}
\cr} \eqno(4.34)$$
 \par Taking the cross product of ${\bf f}$ with the first of $(4.33)$,
one obtains
$$ V^0 ({\bf f} \times {\bf H}) = {\bf V} ({\bf f}\cdot {\bf D}) -
{\bf D} ({\bf f}\cdot {\bf V}).$$
For $V^0 \neq 0$, one may substitute this into the second of
$(4.34)$.  The $f^{05}{\bf D}$ term, with the help of the second of
$(4.33)$, cancels, and we are  left with
$$ {\bf S} = {S^0 \over V^0} {\bf V}.$$
It then follows that $S^\mu = {S^0 \over V^0} V^\mu<$, i.e., $V^\mu$
is proportional to the Poynting vector.  For the case $V^0 = 0$, the
second and third of $(4.33)$ imply that 
$$ {\bf V}\cdot {\bf f} = {\bf V} \cdot {\bf H} =0,$$
i.e., if ${\bf V} \neq 0$ (the case $V^\mu =0$ is exceptional in the
eikonal approximation), it must be proportional to ${\bf f} \times
{\bf H}$. From the first of $(4.33)$, we see that ${\bf V} \times {\bf
D} =0$, and if ${\bf D} \neq 0$, it must be proportional to ${\bf V}$.
The space part of $S^\mu$, from the second of $(51)$ is then
proportional to ${\bf V}$. Under these conditions, the time part of
$S^\mu$ vanishes, and therefore we again obtain the result that
$V^\mu$ is proportional to $S^\mu$. If ${\bf D} = 0$, then $S^0 =0$
and, since ${\bf V}$ is proportional to $ {\bf f} \times {\bf H}$, it
again follows that $V^\mu$ is proportional to $S^\mu$.   
\par From this point, one may follow the same procedure used in the case of
Maxwell's electromagnetism [13] to obtain the
Hamiltonian flow corresponding to the admissible modes. The Lagrangian
associated with the Hamiltonian $(4.29)$ satisfies a minimal principle,
from which it follows that the Hamiltonian flow is geodesic on this manifold. 
 Replacing $f^{\beta
5}$ in $(4.27)$ by $({\cal E}^{-1})^\beta\, _\gamma n^{\gamma 5}$,
one obtains
$$ \bigl\{\delta^\alpha\,_\gamma - M^\alpha\,_{\gamma\mu\nu} p^\mu
p^\nu \bigr\}n^{\gamma 5} = 0, \eqno(4.35)$$
where
$$ M^\alpha\,_{\gamma\mu}\,^\nu = \delta^\alpha\,_\mu ({\cal
E}^{-1})^\nu\,_\gamma - \delta_\mu\,^\nu  ({\cal
E}^{-1})^\alpha\,_\gamma . \eqno(4.36)$$
The condition $(4.23)$ implies that the solutions can lie only in the
hyperplane orthogonal to $p^\sigma$.  The projection of the matrix 
$ M^{\alpha \nu}_{\gamma\mu}p^\mu p_\nu$ (in the indices $\alpha,
\gamma$) into the three dimensional subspace orthogonal to $p^\sigma$
is symmetric, and can therefore be diagonalized by an orthogonal (or
pseudo-orthogonal) transformation in three dimensions.  In fact, the Gauss law
and a gauge condition restrict the polarization degrees of freedom to
three [10] (the eikonal approximation is far from the zero mode,
which corresponds to the Maxwell limit, for which only two
polarizations survive) and hence one finds three geodesics.
\par For $p^\sigma$ timelike, one can choose a Lorentz frame in which
the eigenvalue condition has the form
$$ \bigl\{\delta ^\alpha\,_\gamma - M'^\alpha_{\gamma 00}(p'^0)^2
\bigr\}n'^{\gamma 5} = 0, \eqno(4.37)$$   
 and for $p^\sigma$ spacelike,
$$ \bigl\{\delta ^\alpha\,_\gamma - M''^\alpha_{\gamma 33}(p''^3)^2
\bigr\}n''^{\gamma 5} = 0. \eqno(4.38)$$
 In each of these cases, the matrix can be diagonalized under the little group
acting in the space orthogonal to $p^\sigma$ (leaving it invariant).  For the
 lightlike case, up to a rotation, $p^\sigma$ has the form $(p,0,0, p)$.  The
remaining matrix may then be diagonalized under SO(2) rotations, to
obtain just two geodesics, corresponding to the polarization states of
a massless Maxwell-like theory.  This special limiting case will be
investigated in detail elsewhere.
\par With the same procedure as applied to the Maxwell case [13] one finds the
symplectic structure of the flow of matter in space time.
\par It has been shown by Kline and Kay [13], 
 that for the three
dimensional Maxwell case, the Hamilton equations resulting from the
eikonal coincide with the geodesic flow generated by the resulting
metric (recall that the direction of momentum associated with both
eigenstates is the same); a similar proof can be applied to the 4D
cases we have studied here.
\par The study of the five dimensional wave equations we have carried
out above provide an interesting example of the construction of an
analog gravity.  We have emphasized the important special case where
the dielectric tensor in the Maxwell part of the constitutive
equations is trivial, and sufficient curvature is introduced through
coupling to the fifth component alone.  We now turn to a study of a
quantum equation which results in gravitational physics in the ray
approximation, providing a quantum theory which may underlie the
observed classical gravitational fields.  In a later section, we give
a mechanism, following Nelson[1], based on correlations in relativistic
Brownian motion, providing an interpretation for the structure of this 
equation.
\par  One could think of applying the eikonal method to a Schr\"odinger
equation in a medium which is not isotropic,
for example, in a crystal with shear forces [35], with locally
varying band structure (as in a crystal under nonuniform stress, or near the
boundaries or impurities); the second order derivatives in the
Schr\"odinger equation then appear multiplied by a ``mass matrix.''
\par   The
rays are directly associated with the (probability) flow of particles. The
eikonal eigenvalue condition is one dimensional in this case, since
the field is scalar.  For an analog of this structure (corresponding,
 for example, to
a distribution of events periodic in both space and time) in four
dimensions described by a relativistically covariant
 equation of Stueckelberg-Schr\"odinger type, the metric one obtains is a
spacetime metric, and the geodesic flow is that of
the quantum probability for the spacetime events (matter) described by
the Stueckelberg wave function.  We propose a study of the equation
$$ i{\partial \over \partial \tau} \psi_\tau (x)= -{1 \over 2M}\partial^\mu
g_{\mu \nu}(x) \partial^\nu \psi_\tau(x), \eqno(4.39)$$ 
   where $g_{\mu \nu}(x)$ is assumed to be symmetric, and is somewhat
analogous to a gauge field. We shall refer to it as a {\it tensor gauge
field}. We assume no explicity $\tau$ dependence
in  $g_{\mu \nu}(x)$ in this work. The Schr\"odinger
 current (satisfying the five dimensional conservation law $(2.6)$)is then
$$ j_\tau (x)_\nu = -{i\over 2M} \bigl( {\psi_\tau}^* g_{\mu \nu}
\partial^\mu \psi_\tau - \psi_\tau g_{\mu \nu} \partial^\mu 
{\psi_\tau}^* \bigr) . \eqno(4.40)$$
In the eikonal approximation, for which the frequency associated with
$\tau$ (essentially the total mass of the system [21]) is large,
we assume a form for the solution 
$$ \psi \sim e^{-i(\kappa\tau - \sqrt{\kappa}S)}, $$
where $S$ is the eikonal phase.
\par One obtains the condition, for $\kappa \rightarrow \infty$, 
$$  {1 \over 2M}g^{\mu \nu} p_\mu p_\nu = 1 , \eqno(4.41)$$
where
  $$ p_\mu = \partial_\mu S $$
analogous to the Fresnel surface condition $(4.29)$ for the optical
case. 
\par We define 
$$ K = {p_\mu p_\nu \over 2M} g^{\mu\nu} $$
as the generator of motion in the corresponding classical dynamics.
 It is clear that  $\partial K/ \partial p_\mu$ is in the direction
of $j^\mu_\tau$. This implies that $K$ is the operator of evolution
for the dynamical flow of particles, corresponding to the rays. It
follows from the Hamilton equations that the
flow is geodesic, where $g_{\mu \nu}$ is the metric for this manifold.
\par As in the discussion of the conformal case, we see that
$$ {\dot x}^\mu = {\partial K \over \partial p_\mu} = {g^{\mu \nu} p_\nu \over
M},$$
so that 
           $$ p_\nu = M g_{\mu\nu} {\dot x}^\mu.$$
It then follows that 
$$ K = {1 \over 2}M {\dot x}^\lambda {\dot x}^\sigma g_{\lambda
\sigma}.$$
If we assume that in the flat asymptotic limit the particle is
on-shell, since $K$ conserved, with the value $K= -{M \over
2}$ everywhere, we would obtain
$$ d\tau^2 = -g_{\lambda\sigma} dx^\lambda dx^\sigma. $$
If we identify the world time interval $d\tau$ with a measure of
length (as in free fall), then we can understand $g_{\mu\nu}$ as a
 metric on a psuedo-Riemannian manifold.  As for the conformal
case, however, the function $p_\nu$ induces canonical translations on $x^\nu$.
\par We emphasize that $(4.39)$ is an equation on flat Minkowski space,
and the tensor gauge function $g_{\mu\nu}$ is given as a second rank
  tensor under the Lorentz group.  It is only in the eikonal
  approximation that one finds geodesic flow governed by $g_{\mu\nu}$
  and its affine associated connection with which the manifold is
naturally lifted to a curved space, and the structure may become
  covariant under
  local diffeomorphisms (general covariance). The equivalence
  principle then appears as a characterization of the
  tangent space of this new manifold.
  \par With this equation, one can study a quantum theory which
underlies classical gravity and, in particular study the quantum
behavior in the neighborhood of singularities of the metric for which
smooth eikonals may not exist.  In the case that the eikonal
approximation is valid, the ray approximation provides the geodesics
of the corresponding gravitational field, and assures that the
probability flow is along the geodesic.
\bigskip
\noindent{\bf V. An Interpretation Based on Relativistic Brownian
 Motion}
\smallskip
\par In this section, we shall discuss relativistic Brownian
motion, and follow Nelson for the construction of the
 Stueckelberg-Schr\"odinger equation with tensor gauge coupling.  In
this framework, the tensor coupling arises from correlations between
spacetime dimensions in the underlying Brownian processes. In the following, we
pose and solve some of the difficulties in achieving a definition of 
relativistic Brownian motion.
  \par Brownian motion, thought of as a series of ``jumps'' of a
particle along its path, necessarily involves an ordered sequence.  In
the nonrelativistic theory, this ordering is naturally provided by the
Newtonian time parameter.  In a relativistic framework, the Einstein
time $t$ does not provide a suitable parameter. If we contemplate
jumps in spacetime, to accommodate a covariant formulation, a possible
spacelike interval between two jumps may appear in two orderings in
different Lorentz frames. The introduction of proper time as a
parameter for the relativistic Brownian process (RBP) is not possible
since the second order correlations in the simplest case (i.e. for an
 isotropic homogeneous process with a diffusion constant $\sigma^2$ )
 have the form, for each $\mu$,
 $$E(\Delta x^\mu \Delta x^\mu)=2 \sigma^2 \Delta s \eqno(5.1)
$$
for each $\mu$; however, summing over $\mu$,
$$ E(\Delta x_\mu \Delta x^\mu)\equiv \Delta s^2 \propto
 \Delta s,
 \eqno(5.2)$$
where the first equality is by the definition of proper time and the
 second equality is due to the Brownian property expressed in
 Eq.(5.1).
 There is an obvious contradiction. We therefore adopt the invariant
parameter $\tau$ as the dynamical variable for the Brownian process,
 first
 suggested by Stueckelberg [6]. 
\par  The interpretation of an event
 going backwards in time as the antiparticle was given first by
 Stueckelberg and was later used by Feynman; it is now an accepted
 concept. In Feynman's perturbative formulation of quantum
 electrodynamics [6], pair annihilation and creation occurs at points
 which are sharp vertices at the transitions.  However, on a  smooth
 worldline describing the pair
 annihilation process, as described by Stueckelberg [6], there are segments in
 which the event goes faster than light speed (either forward or backward in
 $t$).  In the 
 formulation of our
 relativistic Brownian process, such sectors appear to play an
 important role; the occurrence of such states of
 motion is dictated by the demand of achieving a Lorentz invariant operator
 (more explicitly, the
 d'Alembertian) in the relativistic diffusion equation.     
\par A second fundamental difficulty in formulating a covariant theory
of Brownian motion lies in the form of the correlation function of the
 random variables of spacetime. The correlation function for the usual
 isotropic non-relativistic Wiener (.e.g. [1]) process is given by 
$$dw^i(t) dw^j(t')= \sigma^2\delta^{ij}\delta(t-t'), $$
for $i,j=1,2,3 $,
where  $dw^i$ corresponds to the Brownian random part of the
Langevin evolution ($\beta^i$ corresponds to a smooth drift)
  $$ dx^i = \beta^i dt + dw^i \eqno(5.3)$$
 A straightforward covariant generalization to the relativistic case is
$$dw_\mu(\tau) dw_\nu(\tau')=
   \alpha^2 \eta_{\mu\nu} \delta(\tau-\tau') \eqno(5.4)$$   
where  $\eta^{\mu \nu}={\rm diag}(-1,1,1,1)$ is the Minkowski
 metric. It then follows that $dw_0(\tau)dw_0(\tau)<0,$
which is impossible. Let us consider, however, a process which is
physically restricted to only to spacelike or timelike jumps. One may
argue that Brownian
 motion in spacetime should be a generalization of the
non-relativistic
 problem,
constructed by observing the non-relativistic process from a moving
frame according to the transformation laws of special relativity (one
could, alternatively, argue that as an idealization of a collision
model, the ``jumps'' should be timelike; we shall consider both
possibilities in the following).
Hence the process taking place in space in the
non-relativistic theory would be replaced by a spacetime process
in which the Brownian jumps are spacelike.  The pure time
(negative) self-correlation does therefore not occur. In order to
meet this requirement, we shall use a coordinatization in terms of
generalized polar coordinates which assures that all jumps are
spacelike. A corresponding distribution for such a relativistic Brownian
probability density could be, for example,  of the form $e^{-{\mu^2\over a
d\tau}}$, where $\mu$ is the invariant spacelike interval of the
jump. This is a straightforward generalization of the standard
Brownian process in 3D, which is generated by a probability
density of the form $e^{-r^2 \over a dt}$, where r is the rotation
invariant (i.e. the vector length)and $a$ is proportional to the
diffusion constant. We shall refer to this function as the
relativistic Gaussian.
\par  As we shall see, a Brownian
motion based on purely spacelike jumps does not, however, yield the
correct form for an invariant diffusion process.  We must therefore
consider the possibility as well that, in the framework of
relativistic dynamics, there are timelike jumps.  The corresponding
distribution would be expected to be of the form
$e^{-{\sigma^2\over b d\tau}}$,
where $\sigma$ is the invariant interval for the timelike jumps, and $b$ is
some constant.
 By suitably
weighting the occurrence of the spacelike process (which we take for
the present discussion to be ``physical'', since its nonrelativistic
limit coincides with
the usual Brownian motion) and an analytic continuation of the
timelike process, we show that
one indeed obtains a Lorentz invariant Fokker-Planck equation in
which the d'Alembert operator appears in place of the Laplace
operator of the 3D Fokker-Planck equation\footnote{${}^3$}{The d'Alembert
operator alone does not, as an evolution kernel, assure that an 
initial positive
density remains positive; additional conditions must be imposed. 
We thank Phillip Pearle for a discussion of this point.}. 
One may, alternatively,
consider the timelike process as ``physical''(as might emerge from a
 microscopic model with scattering) and analytically
continue the spacelike (``unphysical'') process to achieve a
d'Alembert operator with opposite sign.
\bigskip
\noindent{\bf 5.1  Brownian motion in 1+1 dimensions}
\bigskip
\par We consider a Brownian path in $1+1$ dimensions generated by a
 stochastic differential (analogue to the Langevin equation
 and Smoluchowsky process [36]), of the form
$$ dx^\mu (\tau) =  \beta^\mu(x(\tau))d\tau + dw^\mu (\tau),
 \eqno(5.5)$$
where $dw$ is a random process which is a relativistic generalization
of the Wiener process, and
$\beta^\mu$
 is a smooth deterministic field (the drift).
\par We start by considering the second order term in the series
expansion of a function of position of the particle on the world line,
$f(x^\mu(\tau) + \Delta x^\mu)$,
involving the operator
$${\cal O}={\Delta x}^{\mu}{\Delta x}^{\nu}{{\partial} \over
 {\partial}{x}^\mu}{\partial \over \partial {x}^\nu}.
 \eqno(5.6)$$
We have remarked that one of the difficulties in describing Brownian
motion in spacetime is the possible occurrence of a negative value for
the second moment of some component of the Lorentz four vector random
variable. If the Brownian jump is timelike, or spacelike, however, the
components of the four vector are not independent, but must satisfy
the timelike or spacelike constraint.  Such constraints can be
realized by using parameterizations for the jumps in which they
are restricted geometrically to be timelike or spacelike.
  We now separate the random jumps into space-like jumps and time-like
 jumps
accordingly, i.e., for the spacelike jumps,
$$ \Delta w^1=\pm \mu \cosh{\alpha} \,\, , \,\, \Delta w^0 =\mu
 \sinh{\alpha}\eqno(5.7)$$
and for the timelike jumps,
$$ \Delta w^1=\sigma \sinh{\alpha} \,\, , \,\, \Delta
w^0 =\pm
 \sigma \cosh{\alpha} \eqno(5.8)$$
Here we assume that the two sectors have the same distribution
on the hyperbolic variable. We furthermore assume that $\mu$,$\sigma$
are generated by a relativistic Gaussian distribution, working in a
Lorentz frame where the $\alpha$ distribution is assumed to be
independent
 of $\mu,\sigma$ and is uniformly distributed on the
restricted interval  $[-L,L]$ (see discussion below) where $L$ is arbitrary
large. Therefore, in this frame $<\Delta w^\mu>$ is zero (this is true in
all
 frames; see discussion in Section 5) and we pick a
normalization such that (for any component) $<\Delta w^n> \propto
\Delta
 \tau^{n\over 2}$
so to first order in $\Delta \tau$ the
 contribution to $<\cal O>$ comes only from $<\Delta w^\mu \Delta w^\nu>$.
\par  For a particle experiencing space-like
jumps only,  the operator ${\cal O}$ takes the following
 form:
$$ {\cal O}_{spacelike}=
 \mu^2[{\cosh^2}{\alpha}{\partial^2
 \over \partial
x^2}+2\sinh\alpha \cosh\alpha {\partial^2 \over {\partial x \partial
t}}+
\sinh^2\alpha{\partial^2 \over \partial t^2}] \eqno(5.9)$$
If the particle undergoes time-like jumps only the operator
${\cal O}$ takes the form:
$$ {\cal O}_{timelike}= \sigma^2 [\sinh^2
\alpha{\partial^2
 \over \partial^2_x}+2\sinh\alpha \cosh\alpha {\partial^2 \over {\partial_x
\partial_t}}
+\cosh^2\alpha {\partial^2 \over \partial t^2}] \eqno(5.10) $$

\par  Since $\mu,\sigma$ and $\alpha$ are random processes, the
average value
 of
 the operator ${\cal O}$ is the sum of the two averages of
 Eq.(5.9)
 and Eq.(5.10).
 A difference between these two averages, leading to the d'Alembert
 operator can only be obtained by considering the analytic continuation of
the timelike process to the spacelike domain, choosing
$\mu^2=-\sigma^2$. 
\par This procedure is analogous to the effect,
well-known in relativistic quantum scattering theory, of a
physical process in the crossed ($t$)channel on the observed
process in the direct ($s$) channel.  For example, in the LSZ
formulation of relativistic scattering in quantum field theory
 (e.g.,[37]), a
creation operator in the ``in'' state may be moved to the left in the
vacuum expectation value expression for the $S$-matrix, and an
annihilation operator for the ``out'' state may be moved to the
right.  The resulting amplitude, identical to the original one in
value, represents a process that is unphysical; its total
``energy'' (the sum of four-momenta squared) now has the wrong
sign.  Assuming that the $S$-matrix is an analytic function, one
may then analytically continue the energy-momentum variables to
obtain the correct sign for the physical process in the new
channel.  Although we are dealing with an apparently classical
process, as Nelson has shown, the Brownian motion problem gives
rise to a Schr\"odinger equation, and therefore contains
properties of the differential equations of the quantum theory. We
thus see the remarkable fact that one must take into account the
physical effect of the analytic continuation of processes
occurring in a non-physical,
 in this case timelike, domain, on the total observed behavior of the
 system.
\par  In the timelike case, the velocity of the particle
 $\Delta w^1/
\Delta w^0 \leq 1$. We shall here use the dynamical association of
coordinate increments with energy and momentum
$$E=M{\Delta w^0\over \Delta \tau}\,\,\,\,\,\,\,
 p=M{\Delta w^1\over
\Delta \tau},\eqno(5.11)$$
so that
$$\sigma^2 = \bigl({\Delta \tau \over M } \bigr)^2
(E^2-p^2), \eqno(5.12)$$
where $M$ is a parameter of dimension mass associated with the
Brownian particle.
It then follows that  $E^2 - p^2 =\bigl({M \over \Delta \tau}
\bigr)^2\sigma^2 >0$.  For the spacelike
 case, where $p/E >1$, we may consider the transformation to an
 imaginary representation
 $E\rightarrow iE'$ and $p\rightarrow ip'$, for $E', p'$
 real (this transformation is similar to the continuation
 $p\rightarrow ip'$ in nonrelativistic tunnelling, for which the
 analytic continuation appears as an instanton), but $E^2 - p^2
\rightarrow
 p'^2 -E'^2 >0$. In this
 case, we take the analytic continuation such that the magnitude of
$\sigma^2$ remains unchanged, but can be called $-\mu^2$, so that
$E'^2-p'^2=\mu^2$ with $\mu$ imaginary. The spacelike contributions
are therefore
 obtained in this mapping by $E,p\rightarrow iE,ip$ and $\sigma
\rightarrow
 i\mu$,
assuring the formation of the d'Alembert operator when the
 timelike and spacelike fluctuations are added with equal weight (this
equality is consistent with the natural assumption, in this case,
of an equal distribution between spacelike and timelike
contributions).  The preservation of the magnitude of the interval
reflects the conservation of a mass-like property which remains,
as an intrinsic property of the particle, for both spacelike and
timelike jumps. As mentioned before, one recalls the role of
analytic continuation in quantum field theory; for the well known Wick
 rotation (e.g.,[38]),
 however,in that case, only the 0-component is analytically continued
and no
 clear direct
physical idea or quantity is associated with it. In the RBP the
identification of the imaginary 4-momentum is dynamical in
origin. It is due to the Lorentz structure of spacetime, which
distinguishes the transitions ${\Delta {\bf x}\over \Delta t}>1$ from
those
 with
${\Delta{\bf x} \over \Delta t}<1$. Though one may object to the association
of $\Delta x^\mu$ with a dynamical momentum (since the instantaneous
derivative ${dx^\mu \over d\tau}$ is not defined for a Brownian
process) the Brownian motion could be understood as an approximation to a
microscopic process, just as it appears in Einstein's work in
1905 [24], where it is assumed that the Brownian motion is
 produced by
collisions. The effective conservation of $E^2-{\bf p}^2$ as a real quantity
in both timelike and spacelike processes suggests that it is a
physical property which preserves its meaning in both sectors. 
\par With these assumptions,  the cross-term in
hyperbolic functions cancels in the sum, which now takes the form
$${\cal O} = \mu^2\bigl[{\partial^2\over \partial^2_x} -
 {\partial^2
\over \partial t^2}\bigr] \eqno(5.13)$$
Taking into account the drift term in
Eq.(6.2.1), one then finds the relativistic Fokker-Planck equation
$$ {\partial \rho(x,\tau)\over \partial \tau} = \bigl\{-{\partial
 \over
 \partial
x^\mu}\beta^\mu + \langle \mu^2 \rangle {\partial \over \partial
x^\mu}
{\partial \over
\partial x_\mu}\bigr\} \rho(x,\tau), \eqno(5.14)$$
where $\partial/\partial x^\mu$ operates on both $\beta^\mu$ and $\rho$.
\par We see that the procedure we have followed, identifying
$\sigma^2= -\mu^2$ and assuming equal weight, permits us to construct
the Lorentz invariant d'Alembertian operator, as required for
obtaining a relativistically covariant diffusion equation.
 \par To see this process in terms of a higher symmetry, let us define
 the invariant $\kappa^2 \equiv E_t^2 -p_t^2\geq 0$ for the timelike
case; our requirement is then that $E_s^2 -p_s^2 =
 -\kappa^2$ for the spacelike case.
  In the framework of a larger
 group that includes $\kappa$ as part of a three vector $(E, \kappa, p)$, the
 relation for the timelike case can be considered in terms of
 the invariant of the subgroup $O(1,2)$, i.e., $E^2 -\kappa^2 -p^2$.  The
 change in sign for the spacelike case yields the invariant
 $E^2+\kappa^2 -p^2$; we designate the corresponding  symmetry (keeping
 the order of $E$ and $p$) as $O(2,1)$.  These two groups may be
 thought of as subgroups of $O(2,2)$, where there exists a
 transformation which changes the sign of the metric of the subgroups
 holding the quantity $\kappa^2$ constant. The kinematic constraints
 we have imposed correspond to setting these invariants to zero (the
 zero interval in the $2+1$ and $1+2$ spaces).
\par The constraint we have placed on the relation of the timelike and
 spacelike invariants derives from the properties of the
 distribution function and the requirement of obtaining the d'Alembert
 operator, i.e, Lorentz covariance of the diffusion equation. It
 appears that in order for the Brownian motion to result in a
 covariant diffusion equation, the distribution function has a higher
 symmetry reflecting the necessary constraints.  The transformations
$E\rightarrow iE'$ and $p\rightarrow ip'$ used above would then
 correspond to analytic continuations from one (subgroup) sector to
 another.   We shall see a similar structure in the $3+1$ case, where
 the groups involved can be identified with the symmetries of the
 $U(1)$ gauge fields associated with the quantum Stueckelberg-Schr\"odinger
equation.
\par We now investigate the Lorentz invariance of the process and the
 correlation functions.
The averaging operations are summations with weights
(probability) assigned to each quantity in the sum. The sums in the
continuum are, of course, expressed by integrals. If we wish to assign a
relativistic Gaussian distribution function then the hyperbolic
angle integration is
 infinite unless we introduce a cutoff.
The question then arises whether our process is invariant or not.
\par We will show that we can use an arbitrary non-invariant (scalar)
 probability
 distribution (for example, a cutoff on the hyperbolic angle) and still
 obtain Lorentz invariant averages,
 using the imaginary representations of the `unphysical jumps'.
For example, $<\Delta w^\mu>$ stands for a summation with a scalar
weight (given by the density) over all the vectors $\Delta w^\mu$, in
 the
 domain. It is therefore a vector. Moreover under the
imaginary representation of spacelike increments relative to the
timelike ones (here we assume the timelike jumps physical), $\Delta
 w^\mu$
 is a simple vector function over all
spacetime which has the following
 form:
$$\Delta w^\mu=\cases{\Delta w'^\mu \,\, \, \,\,& $\Delta w'^\mu$ timelike \cr
    i \Delta w'^\mu \,\, \,\,\,&$ \Delta w'^\mu$  spacelike\cr}. \eqno(5.15)$$
where the $\Delta w'^\mu$ are real.
The quantity $<\Delta w^\mu>$ (formally written as a discrete sum) is
 given therefore by
$$\eqalign{<\Delta w^\mu>&=\sum P(\Delta w) \Delta w^\mu=\sum_{timelike}
P'(\Delta w') \Delta w'^\mu+i\sum_{spacelike}P'(\Delta
 w')
 \Delta w'^\mu\cr
  &=<\Delta w'^\mu>_{\rm timelike}+i<\Delta
w'^\mu>_{\rm spacelike}\cr}\eqno(5.16)$$
where $P(\Delta w)$ (or $P'(\Delta w')$) is the probability(weight) of
having the vector $\Delta w$ (or $\Delta w'$). The two vectors in the
last equality in Eq.(5.16) are just normal Lorentz vectors. If we now
pick a distribution in a given frame for which the average of each of
them (independent of the other) is zero then $<\Delta w^\mu>=0$ is
true in all frames since
 the $0$-vector is Lorentz invariant.
\par Constructing the second correlation, with the assumption of no
 correlation between spacelike jumps and timelike jumps, we find
$$\eqalign{
<\Delta w^\mu \Delta w^\nu> &=\sum P(\Delta w){\Delta
 w^\mu}{\Delta w^\nu}=\sum_{\rm timelike}P'(\Delta w') \Delta w'^\mu
 \Delta w'^\nu+\cr &+i^2\sum_{spacelike}P'(\Delta w') \Delta
 w'^\mu \Delta w'^\nu \cr&=
\sigma^{\mu \nu}_{timelike}-\sigma^{\mu
 \nu}_{spacelike}\cr &\propto
 \eta^{\mu \nu}D \Delta \tau, \cr} \eqno(5.17) $$
where $\sigma^{\mu\nu}$ is the correlation tensor in each case. 
From the definition of $\Delta w'^\mu$ (a four vector) it follows that
$\sigma^{\mu \nu}$ are real Lorentz tensors.
The last equality in Eq.(5.17) is a demand that could be achieved for
the general $1+n$ case, by assuming that in a given frame there is an
invariant Gaussian distribution where the distribution is uniform in
all angles and that there is a cutoff in the hyperbolic angle. The
sum of the two covariant tensors (each a result of
summation on different sectors) is a Lorentz invariant tensor. The
higher
 correlation functions do not
interest us since they are of higher order in $\rho$ and therefore in
$\Delta \tau$ and do not contribute to the Fokker-Plank
 equation.
\par The mapping given in Eq.(5.15) leads necessarily to a
deviation from the standard mathematical formulation of Brownian
motion. There the probability that a particle starting at $x$ at time
$\tau$ ending at $x'$ at time $\tau'$ is equal to the probability that
the particle starts at $x$ at time $\tau$ passing through any possible
intermediate point $x''$ at time $\tau''<\tau'$ and going from there
to the point $x'$ at time $\tau'$ . This property is expressed
 in the Chapman-Kolmagorov equation (e.g. [36]),
$$p(x,\tau,x',\Delta \tau')=\int_{\cal R}p(x,\tau,x'',
\tau'')p(x'', \tau'',x',\tau')d^4x'', \eqno(5.18)$$      
In the relativistic formulation the vector $\Delta w'=x-x'$ could be a
 timelike vector therefore resulting in a real valued vector $\Delta
 w$ according to the mapping in Eq.(5.15) However, the two
 intermediate vectors $\Delta w'_1=x-x''$ and $\Delta w'_2=x''-x$
 could be spacelike, and take the event out of the real manifold into
 a complex valued coordinate. In this case the Chapman-Kolmagorov
 equation does not hold, and the event may be found outside of the
 real manifold.
In order to build a consistent process one must adopt the concept of
`Brownian jumps' which could be a result for example of a process in
which the event (similar to Einstein's original construction)
undergoes collisions and for each collision, or `jump' the mapping in
Eq.(5.15) holds. Therefore at each point in the physical
manifold the event may take any increment spacelike or timelike (with
a possible complex valued contribution to the averages). However,
although the vector leading from the initial point , say $O$, to the
end point, $A$, may be spacelike and therefore be represented as an
imaginary vector it is understood that the event arrives at the real
spacetime point $A$, never physically leaving the real spacetime. This
 structure separates the two manifolds, spacetime which is real and represents
the physical coordinates of the event and a complex space representing
the processes the event undergoes (virtually) going from one point to
another. This structure differs in that sense from the mathematical
formulation due to Wiener and others, but still it can be shown that
the process is invariant on the average under decomposition into
shorter time subprocesses . In other words, we consider the event
starting at some arbitrary point, and going for some time $\Delta
\tau$. We next decompose the time interval into $M$ intervals,
 so that:
$$\sum_{i=1}^M \Delta \tau_i=\Delta \tau$$
We consider then the expression appearing in the Fokker-Plank equation 
$$<\Delta x^\mu \Delta
x^\nu>=<\bigl(\sum_{i=1}^{M}\Delta
 x_i^\mu\bigr)\bigl(\sum_{j=1}^{M}\Delta x_i^\nu\bigr)>.$$
Since we assume that any two non-equal time jumps are not correlated,
 i.e. $<\Delta x_i\Delta x_j=0>$ for $i\neq j$, which leaves only the
 equal time
 averages in the sum,
$$<\Delta x^\mu \Delta x^\nu>=\sum_{i=1}^{M}<\Delta
x_i^\mu\Delta x_i^\nu>=\sigma^2\eta^{\mu \nu}\sum_{i=1}^M \Delta
\tau_i=\sigma^2 
\eta^{\mu \nu}\Delta \tau$$     
where $\sigma^2$ is the diffusion constant and we used
 Eq.(5.17) going from the second piece in the equality to the
 third.
\par The notion of ``jumps'' suggests the consideration of
 discrete processes, which can also be formulated within the
 relativistic framework and leads, under certain assumptions,
 to a covariant Fokker-Plank equation.
 For example let us assume a physical process in which the ``jumps''
 occur in a very ordered way every $\tau_J$ seconds with a very small
 time spread (i.e. a very small probability that a collision occurs
 within a time different
 significantly from $\tau_J$).
Then, averaging the 'jumps' over a period $\tau>>\tau_J$ leads to
$$<\Delta w^\mu \Delta w^\nu>\cong N\sigma^2\eta^{\mu \nu}\tau_J 
N \tau_J<\tau<(N+1)\tau_J $$
This result is due to the fact that under our assumptions during the
 time
 $\tau$, $N$ single `jumps' within separation of each other of
$\tau_J$ occurred. 
The average in Eq.(5.17)does not change when $\tau$ changes
in less then $\tau_J$; however if $\tau_J$ is small then one can
replace Eq.(5.17) with
$$<\Delta w^\mu \Delta w^\nu>\cong \sigma^2\eta^{\mu \nu}\tau$$
Therefore we recover the standard result for Brownian motion. However
 there is one very important difference which is the fact that $\tau$
 can be taken to be finitely small where in the standard Brownian
 process $\tau$ can be actually taken to zero.This implies that higher
 order derivative terms enter into the resulting `diffusion'
 equation. For example for an isotropic homogeneous Gaussian
 distribution there will be additional even order derivative operators
 beyond the second order (d'Alembert) with coefficients $\sigma^n
 {\tau_J}^(2n-1)$ where $n$ is the (even) order of the differential
 operator. Since both $\tau_J$ and $\sigma^2$ are small these
 operators could be neglected in general, though
 there might be special configurations in which their effect may be
 significant.    
In the following we assume that the $\tau_J$ are very small compared
with the macroscopic scale and that the `jumps' are practically
ordered with zero spread, thus the approximation in
Eq.(5.17) is valid and no higher
 order terms are considered.
\bigskip
\noindent{\bf 5.2 Brownian motion in  $3+1$ dimensions}
\smallskip
\par In the $3+1$ case, we again separate the jumps into timelike and
 spacelike types. The spacelike jumps may be parameterized, in a given
 frame,  by
$$\eqalign{
\Delta w^0 &= \mu \sinh{\alpha}\cr
 \Delta w^1 &= \mu\cosh{\alpha}\cos{\phi}\sin{\vartheta}\cr 
  \Delta w^2 &= \mu\cosh{\alpha}\sin{\phi}\sin{\vartheta}\cr
   \Delta w^3 &= \mu \cosh{\alpha}\cos{\vartheta}\cr} 
 \eqno(5.19)$$
 \par We assume  the four variables $\mu, \alpha, \vartheta, \phi$ are
independent random variables. In addition we demand in this frame that
 $\vartheta$ and $\phi$ are uniformly distributed in their ranges
 $(0,\pi)$ and $(0, 2\pi)$, respectively. In this case, we may average
 over the
 trigonometric angles, i.e., $\vartheta$ and $\phi$ and find that:
$$\eqalign{
<{\Delta w^1}^2>_{\phi , \vartheta}&= <{\Delta
 w^2}^2>_{\phi,\vartheta}= <{\Delta w^3}^2>_{\phi , \vartheta}={\mu^2
 \over 3 }{\cosh}^2{\alpha}\cr
   <{\Delta w^0}^2>_{\phi ,\vartheta}&=\mu^2 {\sinh}^2 {\alpha}\cr}
 \eqno(5.20)$$
We may obtain  the averages over the trigonometric angles of the
timelike jumps by replacing everywhere in Eq.(5.20)
$$
 \cosh^2{\alpha} \leftrightarrow \sinh^2{\alpha}
  \,\,\,\, , \,\,\,\,
 \mu^2 \rightarrow \sigma^2$$
to obtain
$$\eqalign{<{\Delta w^1}^2>_{\phi , \vartheta}= <{\Delta
 w^2}^2>_{\phi,\vartheta}&=
 <{\Delta w^3}^2>_{\phi ,\vartheta}={\sigma^2 \over 3
 }{\sinh}^2{\alpha}\cr
& <{\Delta w^0}^2>_{\phi , \vartheta}= \sigma^2{\cosh}^2
 {\alpha},\cr} \eqno(5.21)$$
 where $\sigma$ is a real random variable, the invariant timelike interval.
Assuming, as in the $1+1$ case, that the likelihood of the jumps being
in either the spacelike or (virtual) timelike phases are equal, and
 making an
analytic continuation for which $\sigma^2 \rightarrow -\lambda^2$,
  the total average of the operator ${\cal O}$, including the
 contributions of
the remaining degrees of freedom $\mu,\lambda$ and $\alpha$  is
$$\eqalign{
<\cal O>&=\bigl(<\mu^2>
<{\sinh}^2{\alpha}>-<\lambda^2><{\cosh}^2{\alpha}>\bigr){\partial^2\over
\partial t^2} \cr
 &{1\over 3}
\bigl(<\mu^2><\cosh^2{\alpha}>-<\lambda^2><\sinh^2{\alpha}>
\bigr){\bigtriangleup}\cr}
 \eqno(5.22)$$
If we now insist that the operator $<{\cal O}>$ is invariant under
Lorentz transformations (i.e. the d'Alembertian) we impose the condition
$$\eqalign{
<\mu^2><{\sinh}^2{\alpha}>-<\lambda^2><{\cosh}^2{\alpha}>&= \cr
 -{1\over 3} \bigl(<\mu^2><\cosh^2{\alpha}>&-<\lambda^2><\sinh^2{\alpha}>
\bigr)\cr} \eqno(5.23)$$

Using the fact that $$<\cosh^2{\alpha}>-<\sinh^2{\alpha}>=1$$, and
defining
 $ \gamma \equiv <\sinh^2{\alpha}>$, we find that
$$<\lambda^2>={1+4\gamma \over 3+4\gamma}<\mu^2>
 \eqno(5.24)$$
The Fokker-Planck equation then takes on the same form as in the $1+1$
case, i.e., the form Eq.(5.14).
We remark that for the $1+1$ case, one finds in the corresponding
 expression that the
$3$ in the denominator is replaced by unity, and the coefficients $4$
are
 replaced by
$2$; in this case the requirement reduces to $<\mu^2> =<\lambda^2>$
 and there is no $\gamma$ dependence.

\par We see that in the limit of a uniform distribution in $\alpha$,
for which $\gamma \rightarrow \infty,$
$$<\lambda^2> \rightarrow<\mu^2>.$$                                     
In this case, the relativistic generalization of 
nonrelativistic Gaussian distribution of the form
 $e^{-{{\bf r}^2\over dt}}$
is $e^{-{\mu^2\over d\tau}}$, which is Lorentz invariant.
\par The limiting case $\gamma \rightarrow 0$ corresponds to a
stochastic process in which in the spacelike case there are no
fluctuations in time, i.e., the process is that of a nonrelativistic
Brownian motion.  For the timelike case (recall that we have assumed
the same distribution function over the hyperbolic variable) this
limit implies that the fluctuations are entirely in the time
direction.  The limit $\gamma \rightarrow \infty$ is Lorentz
invariant,
 but
the limit $\gamma \rightarrow 0$ can clearly be true only in a
particular frame.
\bigskip
\noindent{\bf 5.3  The Markov Relation and the 4D Gaussian Process}
 \smallskip
 \par In developing the previous ideas leading to the formulation of a RBP, we
 assumed that the probability distribution is consistent with the
 Markov property
expressed in the Chapman-Kolmagorov equation. 
However, for the relativistic Gaussian it is not clear whether
 Eq.(5.18) holds. Therefore we now consider an alternative
 process, using the ideas
 developed above, resulting eventually in
the Klein-Gordon equation. Let us consider a 2D Gaussian process generated by a
distribution of the form :
$$p(w,d\tau)={1 \over  2 \pi D d\tau }
 exp({-{\Delta w_0}^2-{\Delta w_1}^2 \over 2 D d\tau})
\eqno(5.25)$$
This distribution corresponds to a Markov process, a standard
normalized Wiener process, where $D$ is the diffusion constant. We now use the
coordinate representation given in Eq.(5.7) and
Eq.(5.8)
 for the timelike and
spacelike sectors to transform the distribution function in
Eq.(5.25)
 in both
sectors to (use $\mu^2$ in both cases):
$${1 \over  2 \pi D d \tau } exp({-\mu^2 \cosh2\alpha
\over
 2 D d\tau})
\eqno(5.26)$$
where timelike `jumps' are physical and the measure for both sectors
 is $\mu
 d\mu d\alpha$ 
 Then,  using Eq.(5.15), we get for the
combination of the timelike and spacelike contributions (with the
 appropriate
 sign)
of the averages, say, ${\Delta w_0}^2$,
$$\eqalign{
<\Delta {w_0}^2>&= {1 \over  2 \pi D d \tau
}\int_0^{\infty}\int_{-\infty}^{\infty}\mu^3 \cr &exp({-\mu^2 \cosh2\alpha \over 2 D
d\tau})d\mu d\alpha =
{1 \over \pi}D d\tau\cr} \eqno(5.27)$$
where we integrated
over $\mu$ first, using
$$\int_0^{\infty}\mu^n exp(-a \mu^2)={\Gamma({ n+1\over
2})\over 2 a^{(n+1)/2}} \eqno(5.28)$$
 and then integrated over $\alpha$ using
$$I_2\equiv \int_{-\infty}^{\infty}{d\alpha \over
 \cosh^2
 2\alpha}=1 \eqno(5.29)$$
 In a similar way one finds that (using Eq.(5.15)) leading to
 the
 negative sign)
$$<{\Delta {w_1}}^2>=-{1 \over \pi}D d\tau \eqno(5.30)$$
Since the probability distribution Eq.(6.5.2) is symmetric in
 ${\Delta w_i}$ in each
sector \break $<\Delta w_0 \Delta w_1>=0$ as well as the first
moments.
 Therefore we
get in this particular frame a d'Alembertian. However, following the
methods used above, we see that it is an invariant result in all Lorentz frames
 (though in other frames
the distribution may not appear to be Gaussian).
\par Next we consider the application of the 4D form of Eq.(5.25)
$$p(w,d\tau)={1 \over  4 \pi^2 D^2 (d\tau)^2 }
 exp({-{\Delta w_0}^2-{\Delta w_1}^2-{\Delta w_2}^2-{\Delta w_3}^2
\over 2 D d\tau}) \eqno(5.31)$$ 
with measure $\mu^3d\mu \cosh^2\alpha \sin\theta d\theta d\alpha d\phi$
for the spacelike sector and $\mu^3d\mu \sinh^2\alpha \sin\theta d\theta
d\alpha d\phi$
 for the timelike sector. 
\par However, now calculating $<{\Delta w_0}^2>$ for the
timelike case, after averaging over the spatial angles $\theta$ and
 $\phi$
 we find,
using Eq.(5.19),
$$<\Delta {w_0}^2>={1 \over \pi D^2 (d\tau)^2
}\int_0^{\infty}\int_{-\infty}^{\infty}\mu^5exp({-\mu^2 \cosh2\alpha \over 2 D
d\tau})\cosh^2\alpha \sinh^2\alpha d\mu d\alpha
\eqno(5.32)$$
 and for the spacelike
case we get the same result since the spacelike parametrization of
 ${\Delta w_0}^2$
is proportional to $\sinh^2\alpha$ and the spacelike volume element is
 proportional
to $\cosh^2\alpha$. Therefore if we use Eq.(5.15), adding the
 contribution of the
two sectors one obtains a complete cancellation to zero. In order to
avoid
 this we
extend Eq.(5.15) to the form
$$\eqalign{
\Delta w^\mu&=\Delta w'^\mu \,\, , \,\, \Delta w'^\mu \,\,{\rm
timelike}\cr
 \Delta w^\mu&=i\lambda \Delta w'^\mu \,\, ,\,\, \Delta w'^\mu \,\, {\rm
spacelike} \cr}\eqno(5.33)$$

Before completing the calculation, we discuss the inclusion of the factor
$\lambda$, in Eq.(5.33). Let us consider a classical (i.e.
 non-stochastic) event with
a given value $m^2 \equiv \Delta w_\mu \Delta w^\mu$, moving in a timelike
direction. It then changes its state of motion and starts moving in a spacelike
direction; according to Eq.(5.33) $m^2$ changes into $\lambda^2 m^2$.
Moreover, though the event may move according to a Gaussian
distribution
 which makes
no distinction between timelike and spacelike motions, the outcome of
this
 motion as
represented by the $\Delta w$, in Eq.(5.15);
Eq.(5.33) does
 distinguish the two
phases of motion. We shall see that a specific value of $\lambda$ is
required
 for the realization of the Fokker-Planck equation.
\par The $w$ manifold is complex and it is a function of the motion on the real
manifold $w'$. Our macroscopic (physical) equations are written on the
 real
 plane of
the $w$ manifold. One can then visualize the flow of an event in
spacetime
 similar
to a motion of a particle in a cloud chamber. There as the particle moves the
gas condenses, therefore the particle leaves a track. The track itself is not
the particle but a result of the actual motion of the particle and its
 interaction
with the gas in the cloud chamber.  The track in the cloud chamber is
 analogous to the complex representation we use for the `jumps'.  

We calculate first the expectation $<{\Delta w_0}^2>$ which is the
 total
 expectation,
summed over the timelike and spacelike sectors. Averaging over the
spherical
 angles
$\theta,\varphi$ we get using Eq.(5.32) and Eq.(5.33),
$$\eqalign{
<\Delta {w_0}^2>&={1 -\lambda^2\over \pi} D^2 (d\tau)^2
\int_0^{\infty}\int_{-\infty}^{\infty}\mu^5exp({-\mu^2 \cosh2\alpha \over 2 D
d\tau})\cosh^2\alpha \sinh^2\alpha d\mu d\alpha \cr
&= {8 (1-\lambda^2)\over \pi}D
d\tau \int_{-\infty}^{\infty}{\cosh^2\alpha \sinh^2\alpha \over \cosh^3
2\alpha}d\alpha \cr }\eqno(5.34)$$
where Eq.(5.28) was used in the $\mu$ integration leading to
 the last
 equality
in Eq.(5.34).
\par  Using
$$ \cosh^2\alpha = {1\over 2}(\cosh 2\alpha+1)
\sinh^2 \alpha={1\over2}(\cosh2\alpha-1)\eqno(6.5.12)$$
in Eq.(5.34) and integrating over $\alpha$ we get
$$<\Delta {w_0}^2>={(1-\lambda^2)\over \pi}D d\tau 
{\pi \over 2} \eqno(5.36)$$
where we used
$$\eqalign{I_1&\equiv \int_{-\infty}^{\infty}{d\alpha \over
 \cosh 2\alpha}={\pi \over 2 }\cr
 I_3&\equiv \int_{-\infty}^{\infty}{d\alpha \over \cosh^3
 2\alpha}={\pi
 \over 4}\cr} \eqno(5.37)$$
\par We now calculate the expectation of $<{\Delta w_1}^2>$. Averaging over
 the spherical
angles $\theta,\varphi$ we get, using Eq.(5.32) and Eq.(5.33),
$$\eqalign{
<\Delta {w_1}^2>&={1\over 3 \pi} D^2 (d\tau)^2
\int_0^{\infty}\int_{-\infty}^{\infty}\mu^5exp({-\mu^2 \cosh2\alpha \over 2 D
d\tau})(\sinh^4 \alpha-\lambda^2\cosh^4\alpha)d\mu d\alpha =
\cr
&= {8 \over 3 \pi}D
d\tau \int_{-\infty}^{\infty}{(\sinh^4\alpha-\lambda^2\cosh^4\alpha)
 \over \cosh^3
2\alpha}d\alpha \cr}\eqno(5.38)$$
\par Using Eqs.(5.35), (5.29)and (5.37), We
 get after
 integration over $\alpha$,
$$<\Delta {w_1}^2>={D d\tau \over \pi}{1 \over
3}[(1-\lambda^2
){3\pi \over
2}-4(1+\lambda^2)]$$                                  
In order to obtain the d'Alembertian we insist that $<\Delta {w_1}^2>=-<\Delta
{w_0}^2>$, which leads to
$$ \lambda^2={3\pi-4 \over 3\pi+4} \eqno(5.39)$$
Finally, substituting (for example) Eq.(5.39) in
 Eq.(5.36)
 we find that
$$<\Delta w^\mu \Delta w^\nu>=\eta^{\mu \nu}{4D \over
 3\pi +4}d\tau= \eta^{\mu \nu} \breve{D}d\tau \eqno(5.40)$$
where $\breve{D}$, is the actual effective diffusion constant defined by
$$ \breve{D}\equiv {4D \over 3 \pi +4} \eqno(5.41)$$
\bigskip
\noindent {\bf VI. Discussion and Conclusions}
\smallskip
\par We have discussed in the first section the properties of a
classical relativistic charged particle in the framework of a
consistent relativistically covariant classical dynamics.
Such a theory, without a constraint relating the particle variables
$E$ and ${\bf p}$ admits variations of the particle mass from the
``mass-shell'' (for which an interval of proper time is equal to the
corresponding interval of the universal world-time); this mass
variation plays an important role in governing the evolution of the
system, and may in fact, macroscopically stabilize the spacetime orbit
in the presence of the highly singular self-interaction.  The work of
Gupta and Padmanabhan shows that the radiation reaction terms in the
Lorentz force of the self-interacting system have a geometrical
interpretation. In view of this result, one is motivated to study a
possible connection between such dynamical systems and gravity.
\par  In the next section, we showed that the $a_5$ field,
primarily reesponsible for driving the particle off-shell, can indeed be
absorbed in an effective conformal metric structure for the motion on a
manifold. We showed that one can compare this form to the
Robertson-Walker-Friedmann model, and, through the field equations,
obtain a relation for the matter density and understand how, in such a
world, the Einstein time (in conformal coordinates) and $\tau$
dependence can be constrained to be similar.
\par Following this idea further, we find that the eikonal
approximation to wave equations for the (generalized) propagation of
electromagnetic waves in a medium with non-trivial dielectric tensor
can result in an analog gravity.
\par   A mathematically simpler system with
this property is that of a Stueckelberg-Schr\"odinger equation with a
``tensor gauge'' coupling.  This equation permits one to study a
quantum theory on a flat space, for which the eikonal approximation
yields a system of rays which correspond to the flow of probability
along the geodesics of a manifold with metric determined by the tensor
gauge coupling.  Such a system can be lifted by general covariance to give the
complete structure of general relativity.  One can therefore study, in
this way, the properties of a well-defined canonical quantum theory
for which the eikonal approximation is classical gravity.  The quantum
properties of such a system in the neighborhood of singularities, such
as the black hole horizon, may yield interesting information on
questions such as Bekenstein's conjecture about the spectrum of black
hole radiation. 
\par To understand the structure of a Stueckelberg-Schr\"odinger
equationm of this form, we appealed to Nelson's construction of the
Schr\"oding equation from the process of Brownian motion. To do this,
it was necessary to establish a relativic form of Brownian motion, and
we described and solved the difficulties in achieving such a result. 
\par In particular, we  constructed a relativistic generalization of
 Brownian               
motion, using the invariant world-time, $\tau$, to order the Brownian
 fluctuations, and separated consideration of spacelike and
 timelike            
 jumps to avoid the problems of negative second moments which
 might             
 otherwise follow from the Minkowski signature. Associating
 the Brownian fluctuations with an underlying dynamical process, one may
 think of $\gamma$ discussed in the $3+1$ case as an order parameter,
 where the distribution function (over $\alpha$), associated with the
 velocities, is determined by the temperature of the underlying
 dynamical system (the result for the $1+1$ case is independent of the
 distribution on the hyperbolic variable).  More generally it is
 suggestive to consider the possible thermodynamical effects of the
 `medium' generating the relativistic Brownian fluctuations, following
 similar steps taken by Einstein [Einstein] in his famous work
 and verify whether any physical effect can be
 predicted.
\par  At equilibrium, where $\partial \rho/\partial \tau =0,$ the
 resulting diffusion equation turns into a classical wave 
equation              
 which, in the absence of a drift term $K^\mu$, is the wave equation 
for a massless field.  An exponentially decreasing distribution
 in $\tau$ of the form $\exp{-\kappa \tau}$ would correspond
 to a                  
 Klein-Gordon equation for a particle in a tachyonic state
 (mass                
 squared $-\kappa$), for physical spacelike motion and for physical
 timelike motion to a particle with mass squared $\kappa$.                   
\par Choosing a cutoff in the hyperbolic angle, one finds a covariant
 moment and, therefore, 
covariant differential operators. However the underlying process
 is not invariant; thus one can think of a special frame in which the       
 hyperangular distribution is uniformly distributed around 0.  Boosting 
breaks the symmetry of the hyperangular distribution, but since the
 the averages are tensor quantities the invariance properties are
 conserved, and therefore the Fokker-Plank equation (leading to the
 quantum equation) is invariant.
 This property is also used to construct the 4D Gaussian process.
 \par It was shown that a (Euclidian) Gaussian process with an appropriate 
 (weighted) complex representation for the timelike and spacelike
 random motions can be used to achieve the covariant quantum equation,
 with the assurance that it is Markovian (it is a relativistic
 generalization of the Wiener process). This leads to a `cloud
 chamber-like' picture in which the event as it evolves leaves a track
 (carries a real or an imaginary phase),which is a
 representation of the actual motion, distinguishing the timelike and
 spacelike motion.
\par In the classical Stueckelberg theory the timelike (forward or
 backward) propagation is associated with the standard particle or
 antiparticle interpretation, where spacelike propagation is needed
 whenever one discusses classical pair creation or annihilation (with
 continuous passage from forward to backward motion in time). This
 suggests that the spacelike process may be associated with
 the annihilation and creation of pairs.
Moreover, though the resulting  macroscopic equation(i.e. on the level
 of the Fokker-Planck equation) is local and causal in the spacetime
 variables, the underlying microscopic process(i.e., on the level of
 the Brownian fluctuations) is not. It is however local and causal in
 $\tau$ even at the microscopic level.
 \par  Nelson has shown that non-relativistic Brownian motion can be
 associated with a Schr\"odinger equation. Equipped with the
 procedures we presented here, constructing relativistic Brownian
 motion, Nelson's methods can be generalized. One then can construct
 relativistic equations of Schr\"odinger (Schr\"odinger-Stueckelberg)
 type. The eigenvalue equations for these relativistic forms are also
 Klein-Gordon type equations. Moreover one can also generalize the
 case where the fluctuations are not correlated in different
 directions into the case where correlations exist, as discussed by
 Nelson for three dimensional Riemannian spaces. In this case the
 resulting equation will be a quantum equation with a local tensor
 coupling; as we have pointed out, the eikonal approximation to the
 solutions of such an equation contains the geodesic motion of
 classical general relativity. The medium supporting the Brownian
 motion may be identified with an `ether'  for which the problem of
 local Lorentz symmetry is
 solved. 
 This study opens up several tracks of possible research. Nelson [1],
 discussing the E.P.R. system confronted the fact
 that such a system may be either described by a non-local Markov
 process or a local non-Markov process. The Markov process is simple
 to implement but Nelson was disturbed by the introduction of
 non-local interaction. However, the non-Markov process is very
 difficult to apply. The RPB developed here may bridge the two
 possibilities since an ordered (causal) Markov process in $\tau$ may
 appear to be a non-Markovian (or possibly non-local and certainly non
 causal) process in $t$. For example for the Gaussian process, looking
 for the probability of finding the event changing its spatial
 position $\Delta x$ after $\Delta t$ has passed, one may integrate
 Eq.(5.25) over all $\tau$. This results however, in ${{1
 \over {\Delta x}^2+{\Delta t}^2}}$ which is not integrable and
 therefore can not be normalized. This however is not surprising,
 since the probability of finding the event in $\Delta x$ after
 $\Delta t$ is not well defined (there may be several values of
 $\Delta x$ for a given $\Delta t$). For example the number of
 particles, as in Stueckelberg's original construction
 [6] depends on the trajectory through which the point
 is reached. Defining an appropriate one particle probability
 resulting from the initial process occurring in $\tau$ demands a
 restriction of the sample space before integrating over $\tau$ i.e.,
 using the conditional probability restricted to processes for which
 the event's
 $t$ coordinate is monotonic in $\tau$ (no pairs are created).     
 \par Finally we would like to point out that generating a covariant
 quantum equation through an RBP leads to a possible relation between
 quantum mechanics and gravitation. In the context of this work, the
 metric of gravity can appear as an anisotropy  in the correlations
 that lead to quantum equations for which the ray, or eikonal
 approximation, corresponds to the classical geodesic flow of general
 relativity. It furthermore appears interesting to generalize
 Einstein's famous work on this process introducing thermodynamic
 concepts to the
 resulting geometrical structure of the theory.
\par It emerges that the spacetime metric associated with the eikonal
 geodesic flow, in such a theory, would have its origin in correlations
in the underlying Brownian process.  
\bigskip
\noindent{\it Acknowledgements}
\bigskip
\par One of us (L.P.H.) would like to thank the Institute for Advanced
Study, Princeton, N.J. for partial support, and Steve Adler for his
hospitality, during his visit in the Spring Semester (2003) when much
of this work was done. He also wishes to thank Philip Pearle for
helpful discussions.
 \par We wish to thank C. Piron and F.W. Hehl for
 discussions at an early stage of this work, and H. Goldenberg and N. Erez
 for helpful discussions of propagation of waves in a medium, and also
to thank V. Dvoeglazov for inviting us to write this contribution.
\bigskip
\bigskip                           
\noindent {\it References}
\frenchspacing
\item{1.} E. Nelson, Phys. Rev. {\bf 150} (1966) 1079-1085
see also Edward Nelson, {\it Dynamical Theories of Brownian Motion},
  Princeton University Press, Princeton (1967);      
       {\it Quantum Fluctuations},
 Princeton University Press Princeton (1985) see also Ph. Blanchard,
 Ph. Combe and W. Zheng, {\it Mathematical and Physical Aspects of
 Stochastic Mechanics},
 Springer-Verlag, Heidelberg (1987), for further helpful discussion.        
\item{2.}  A. Gupta and T. Padmanabhan, Phys. Rev. {\bf D57},7241
(1998).
\item{3.} F. Rohrlich, {\it Classical Charged Particles \/}, 
Addison-Wesley, Reading (1965);
 M. Abraham, {\it Theorie der Elektrizit\"at \/}, vol. II,
Springer. Leipzig (1905);
 P.A.M. Dirac, Proc. Roy. Soc.(London) {\bf A167}, 148
(1938);  A.A. Sokolov and I.M. Ternov, {\it Radiation from
Relativistic Electrons\/}, Amer. Inst. of
Phys. Translation Series, New York (1986).
\item{4.} B. Mashoon, Proc. VII Brazilian School of Cosmology and
Gravitation, Editions Fronti\'eres (1994); Phys. Lett. A {\bf 145},
147 (1990); Phys. Rev. A {\bf 47}, 4498 (1993).
\item{5.} C.W. Misner, K.S. Thorne and J.A. Wheeler, {\it
Gravitation\/}, W.H. Freeman, San Francisco (1973).
\item{6.} E.C.G. Stueckelberg, Helv. Phys. Acta {\bf 14}, 322, 588 (1941);
J.S. Schwinger, Phys. Rev. {\bf 82}, 664 (1951); R.P. Feynman,
Rev. Mod. Phys. {\bf 20}, 367 (1948); R.P. Feynman, Phys. Rev. {\bf 80},
 440 (1950). C. Piron and L.P. Horwitz, Helv. Phys. Acta {\bf 46}, 316
(1973), extended this theory to the many-body case throught he
postulate of a universal invariant evolution parameter.
\item{7.} L.P. Horwitz, Found. Phys. {\bf 25}, 39 (1995).
\item{8.} L. Burakovsky, L.P. Horwitz and W.C. Schieve, Phys. Rev. D
{\bf 54}, 4029 (1996).
\item{9.}  O. Oron and L.P. Horwitz, Phys. Lett. {\bf A 280}, 265
(2001); N. Katz, L.P. Horwitz and O. Oron, Discrete Dynamics in Nature
and Society (DDNS), to be published (2004).
\item{10.} D. Saad, L.P. Horwitz and R.I. Arshansky, Found. Phys. {\bf
19}, 1126 (1989);  
 M.C. Land, N. Shnerb and L.P. Horwitz, Jour. Math. Phys. {\bf 36},
3263 (1995);
 N. Shnerb and L.P. Horwitz, Phys. Rev {\bf A48}, 4058
(1993); M.C. Land and L.P. Horwitz, Found. Phys. Lett. {\bf 4}, 61
(1991);
 J.D. Jackson, {\it Classical Electrodynamics}, 2nd edition,
Wiley, N.Y. (1975).
\item{11.}O. Oron and L.P. Horwitz, Foundations of Physics {\bf 33}, 1177 
(2003).
\item{12.} H. Goldstein, {\it Classical Mechanics}, Addison-Wesley,
N.Y. (1950)
\item{13.}M. Kline and I.W. Kay, {\it Electromagnetic Theory and
Geometrical Optics}, John Wiley and Sons, N.Y. (1965).
\item{14.} P. Piwnicki,{\it Geometrical approach to light in
inhomogeneous media}, gr-qc/0201007.
\item{15.} M. Visser, Class. Quant. Grav. {\bf 15}, 1767 (1998);
L.J. Garay, J.R. Anglin, J.I. Cirac and P. Zoller, Phys. Rev. {\bf
A63}, 026311 (2001); Phys. Rev Lett. {\bf 85}, 4643 (2000).
\item{16.}M. Visser, Carlos Barcelo, and Stefan Liberati, Analogue
models of and for gravity gr-qc/0111111 (2001);
 Carlos Barcel\'o and Matt Visser, Int. Jour. Mod. Phys. D {\bf
10}, 799 (2001);
 Carlos Barcel\'o, Stefano Liberati and Matt Visser,
Class. and Quantum Grav. {\bf 18}, 3595 (2001).
\item{17.} U. Leonhardt and P. Piwnicki, Phys. Rev A {\bf 60}, 4301
(1999).
\item{18.} V.A. De Lorenci and R. Klippert, Phys. Rev. D {\bf 65},
064027 (2002). 
\item{19.}  Y.N. Obukhov and F.W. Hehl, Phys. Lett. B 458(1999) 466.  See
also, M. Sch\"onberg, Rivista Brasileira de Fisica {\bf 1}, 91 (1971); A Peres,
Ann. Phys.(NY) {\bf19},279 (1962).
\item{20.} H. Urbantke, J. Math. Phys. {\bf 25} 2321(1984); Acta Phys.
 Austriaca Suppl. XIX,875 (1978).
\item{21.} R. Arshansky and L.P. Horwitz, Jour. Math. Phys. {\bf 30},
66,380 (1989).
\item{22.}M.A. Trump and W.C. Schieve, {\it Classical Relativistic
Many-Body Dynamics,} Kluwer Academic, Dordrecht (1999).
\item{23.} N.D. Birrell and P.C.W. Davies, {\it Quantum Fields in Curved 
Space}, Cambridge University Press, Cambridge (1982).
\item{24.} A.Einstein, Ann. der Physik {\bf 17} 1905
 549-560. See also R.Furth, {\it Investigation of the Theory of
 Brownian Movement}, N.Y, Dutton (1956);
 R.Brown, Philosophical Magazine N.S {\bf 4}, 161-173,(1828);
 R.Brown, Philosophical Magazine N.S {\bf 6},
 161-166,(1829).
\item{25.} L.Morato and L.Viola, J. Math. Phys. {\bf 36} (1995)
 4691-4710.        
\item{27.} O. Oron and L.P. Horwitz, in {\it Gravitation, Cosmology
and Relativity}, ed. V. Dvoeglazov and A. Espinoza Garrido, p. 94-104,
Nova Science, Hauppage (2004).
\item{28.} J.D. Jackson, {\it Classical Electrodynamics}, 2nd edition,
John Wiley and Sons, New York (1975).
\item{29.} A.O. Barut, {\it Electrodynamics and Classical Theory of
Fields and Particles}, Dover, N.Y. (1964).
\item{30.} F. Rohrlich, Found. Phys. {\bf 26}, 1617 (1996);
Phys. Rev.D{\bf 60}, 084017 (1999); Amer. J. Phys. {\bf 68},1109
(2000).  See also,  H. Levine,
 E.J. Moniz and D.H. Sharp, Amer. Jour. Phys. {\bf 45}, 75 (1977);
A.D. Yaghjian, {\it Relativistic Dynamics of a Charged Sphere},
Springer, Berlin (1992).
\item{31.} O. Oron and L.P. Horwitz, {\it Measurability of Classical
 Off-Mass-Shell Effects }, in preparation.
\item{32.} S. Weinberg, {\it Gravitation and Cosmology}, Wiley, New
York (1972); R.M. Wald, {\it General Relativity}, University of Chicago
Press, Chicago (1984).
\item{33.}D. Zerzion, L.P. Horwitz and R. Arshansky,
Jour. Math. Phys. {\bf 32}, 1788(1991).
\item{34.} O. Oron and L.P. Horwitz, Found. Phys. {\bf 33}, 1323
(2003).
\item{35.} H. Brooks, Adv. in Electronics {\bf 7}, 85 (1955);
C. Kittel, {\it Quantum Theory of Solids}, p. 131, John Wiley and
Sons, N.Y. (1963). 
\item{36.} For example, H. Risken, {\it The Fokker-Planck
 Equation\/},Springer, N.Y. (1996). 
\item{37.} M. Kaku, {\it Quantum Field Theory\/ }, Oxford, New York
(1993).
\item{38.} For example, M.E. Peskin and D.V. Schroeder, {\it An
Introduction to Quantum Field Theory\/}, Addison Wesley, Reading
(1995).

\vfill
\bye
\end